\documentclass{pasj01}
\usepackage{multirow}
\usepackage{natbib}
\Received{$\langle$2017 December 08$\rangle$}
\Accepted{$\langle$2018 February 14$\rangle$}
\Published{$\langle$publication date$\rangle$}

\begin{document}
\title{Subaru High-z Exploration of Low-Luminosity Quasars (SHELLQs) III. Star formation properties of the host galaxies at $z \gtrsim 6$ studied with ALMA} 
\author{ Takuma \textsc{Izumi},\altaffilmark{1,}$^{\dag,*}$ 
Masafusa \textsc{Onoue},\altaffilmark{1,2} 
Hikari \textsc{Shirakata},\altaffilmark{3} 
Tohru \textsc{Nagao},\altaffilmark{4} 
Kotaro \textsc{Kohno},\altaffilmark{5,6} 
Yoshiki \textsc{Matsuoka},\altaffilmark{4} 
Masatoshi \textsc{Imanishi},\altaffilmark{1,2} 
Michael A. \textsc{Strauss}, \altaffilmark{7} 
Nobunari \textsc{Kashikawa},\altaffilmark{1,2} 
Andreas \textsc{Schulze},\altaffilmark{1,}$^{\ddag,}$ 
John D. \textsc{Silverman},\altaffilmark{8} 
Seiji \textsc{Fujimoto},\altaffilmark{9} 
Yuichi \textsc{Harikane},\altaffilmark{9} 
Yoshiki \textsc{Toba},\altaffilmark{10} 
Hideki \textsc{Umehata},\altaffilmark{11,5} 
Kouichiro \textsc{Nakanishi},\altaffilmark{1,2} 
Jenny E. \textsc{Greene},\altaffilmark{12} 
Yoichi \textsc{Tamura},\altaffilmark{13} 
Akio \textsc{Taniguchi},\altaffilmark{5} 
Yuki \textsc{Yamaguchi},\altaffilmark{5} 
Tomotsugu \textsc{Goto},\altaffilmark{14} 
Yasuhiro \textsc{Hashimoto},\altaffilmark{15} 
Soh \textsc{Ikarashi},\altaffilmark{16} 
Daisuke \textsc{Iono},\altaffilmark{1,2} 
Kazushi \textsc{Iwasawa},\altaffilmark{17} 
Chien-Hsiu \textsc{Lee},\altaffilmark{1} 
Ryu \textsc{Makiya},\altaffilmark{8,18} 
Takeo \textsc{Minezaki},\altaffilmark{5} 
and Ji-Jia \textsc{Tang}\altaffilmark{9}
}
\altaffiltext{1}{National Astronomical Observatory of Japan, 2-21-1 Osawa, Mitaka, Tokyo 181-8588, Japan }
\altaffiltext{2}{Department of Astronomical Science, Graduate University for Advanced Studies (SOKENDAI), 2-21-1 Osawa, Mitaka, Tokyo 181-8588, Japan } 
\altaffiltext{3}{Department of Cosmosciences, Graduate School of Science, Hokkaido University, N10 W8, Kitaku, Sapporo, 060-0810, Japan} 
\altaffiltext{4}{Research Center for Space and Cosmic Evolution, Ehime University, Matsuyama, Ehime 790-8577, Japan} 
\altaffiltext{5}{Institute of Astronomy, Graduate School of Science, The University of Tokyo, 2-21-1 Osawa, Mitaka, Tokyo 181-0015, Japan} 
\altaffiltext{6}{Research Center for the Early Universe, Graduate School of Science, The University of Tokyo, 7-3-1 Hongo, Bunkyo, Tokyo 113-0033, Japan} 
\altaffiltext{7}{Princeton University Observatory, Peyton Hall, Princeton, NJ 08544, USA} 
\altaffiltext{8}{Kavli Institute for the Physics and Mathematics of the Universe (WPI), The University of Tokyo Institutes for Advanced Study,
The University of Tokyo, Kashiwa, Chiba 277-8583, Japan}
\altaffiltext{9}{Institute for Cosmic Ray Research, The University of Tokyo, Kashiwa, Chiba 277-8582, Japan}
\altaffiltext{10}{Academia Sinica Institute of Astronomy and Astrophysics, P.O. Box 23-141, Taipei 10617, Taiwan}
\altaffiltext{11}{The Open University of Japan, 2-11 Wakaba, Mihama-ku, Chiba 261-8586, Japan} 
\altaffiltext{12}{Department of Astrophysics, Princeton University, Princeton, NJ, USA} 
\altaffiltext{13}{Division of Particle and Astrophysical Science, Graduate School of Science, Nagoya University, Chikusa-ku, Nagoya, Aichi 464-8602, Japan}
\altaffiltext{14}{Institute of Astronomy and Department of Physics, National Tsing Hua University, Hsinchu 30013, Taiwan}
\altaffiltext{15}{Department of Earth Sciences, National Taiwan Normal University, Taipei 11677, Taiwan}
\altaffiltext{16}{Kapteyn Astronomical Institute, University of Groningen, P.O. Box 800, 9700 AV Groningen, Netherlands}
\altaffiltext{17}{ICREA and Institut de Ci\`{e}ncies del Cosmos, Universitat de Barcelona, IEEC-UB, Mart\'{i} i Franqu\`{e}s, 1, E-08028 Barcelona, Spain}
\altaffiltext{18}{Max-Planck-Institut f{\"u}r Astrophysik, Karl-Schwarzschild Str. 1, D-85741 Garching, Germany}
\altaffiltext{$\dag$}{NAOJ Fellow}
\altaffiltext{$\ddag$}{EACOA Fellow}
\email{takuma.izumi@nao.ac.jp}

\maketitle

\begin{abstract}
We present our ALMA Cycle 4 measurements of the [C\,\emissiontype{II}] emission line 
and the underlying far-infrared (FIR) continuum emission from 
four optically low-luminosity ($M_{\rm 1450} > -25$) quasars at 
$z \gtrsim 6$ discovered by the Subaru Hyper Suprime Cam (HSC) survey. 
The [C\,\emissiontype{II}] line and FIR continuum luminosities lie in the ranges 
$L_{\rm [C\,\emissiontype{II}]} = (3.8-10.2) \times 10^8~L_\odot$ 
and $L_{\rm FIR} = (1.2-2.0) \times 10^{11}~L_\odot$, 
which are at least one order of magnitude smaller than those of 
optically-luminous quasars at $z \gtrsim 6$. 
We estimate the star formation rates (SFR) of our targets as $\simeq 23-40~M_\odot ~{\rm yr}^{-1}$. 
Their line and continuum-emitting regions are marginally resolved, 
and found to be comparable in size to those of optically luminous quasars, 
indicating that their SFR or likely gas mass surface densities 
(key controlling parameter of mass accretion) are accordingly different. 
The $L_{\rm [C\,\emissiontype{II}]}/L_{\rm FIR}$ ratios of the hosts, $\simeq (2.2-8.7) \times 10^{-3}$, 
are fully consistent with local star-forming galaxies. 
Using the [C\,\emissiontype{II}] dynamics, 
we derived their dynamical masses within a radius of 1.5--2.5 kpc as 
$\simeq (1.4-8.2) \times 10^{10}~M_\odot$. 
By interpreting these masses as stellar ones, 
we suggest that these faint quasar hosts are on 
or even below the star-forming main sequence at $z \sim 6$, 
i.e., they appear to be transforming into quiescent galaxies. 
This is in contrast to the optically luminous quasars at those redshifts, which show starburst-like properties. 
Finally, we find that the ratios of black hole mass to host galaxy dynamical mass 
of the most of low-luminosity quasars including the HSC ones 
are consistent with the local value. 
The mass ratios of the HSC quasars can be 
reproduced by a semi-analytical model that assumes 
merger-induced black hole-host galaxy evolution. 
\end{abstract}

\section{Introduction}\label{sec1}
Mass accretion onto a supermassive black hole 
(SMBH, with a mass of $M_{\rm BH} \gtrsim 10^6~M_\odot$) 
produces an enormous amount of energy, observable as 
an active galactic nucleus (AGN) or a quasar \citep{1964ApJ...140..796S}. 
SMBHs reside at the centers of massive galaxies, 
and show tight correlations between $M_{\rm BH}$ and the properties of the host galaxies, 
such as bulge stellar mass ($M_{\rm bulge}$) 
and stellar velocity dispersion ($\sigma_*$) in the local universe 
\citep[e.g.,][]{2000ApJ...539L...9F,2003ApJ...589L..21M,2013ARA&A..51..511K}. 
The remarkable similarity between global star formation 
and mass accretion histories \citep[][for a review]{2014ARA&A..52..415M}, 
as well as correlations between luminosities associated 
with AGN and with star formation in luminous systems
\footnote{Note that recent works suggest that this trend is driven 
by a dependence of SFR on the redshift and stellar mass \citep[e.g.,][]{2017ApJ...842...72Y}.} 
\citep[e.g.,][]{2010ApJ...712.1287L,2013ApJ...773....3C}, 
support the rapid growth of SMBHs in tandem with the stellar mass build-up of galaxies. 

Physical mechanisms that may lead to such co-evolutionary scenarios include 
mergers of galaxies and subsequent AGN feedback to regulate star formation in the host. 
Both AGN and star formation may fed by a common supply of the cold interstellar medium (ISM). 
Hydrodynamic simulations based on this framework 
reproduce the observed properties of AGN and star formation 
\citep[e.g.,][]{2005Natur.433..604D,2006ApJS..163....1H,2007ApJ...665..187L}. 
Detections of massive AGN-driven outflows 
\citep[e.g.,][]{2008A&A...491..407N,2012A&A...537A..44A,2012ApJ...746...86G,2012MNRAS.425L..66M,2014A&A...562A..21C,2017arXiv171002525T} 
may also provide an important coupling between the SMBH and its host galaxy. 
Semi-analytic galaxy evolution models 
\citep[][for a review]{2015ARA&A..53...51S} 
predict intense star formation (star formation rate (SFR) reaching $100-1000$ $M_\odot$ yr$^{-1}$) 
and SMBH accretion (accretion rate reaching $10$ $M_\odot$ yr$^{-1}$) 
with very short characteristic time scales, 
on the order of 100 Myr \citep[e.g.,][]{2008ApJS..175..356H,2015MNRAS.449.1470V}, 
particularly at the peak epoch of galaxy formation ($z \sim 2-3$). 

One effective way to further test galaxy evolution models 
is to determine whether co-evolutionary scenarios have arisen 
in the early universe \citep{2012Sci...337..544V,2017PASA...34...22G,2017PASA...34...31V}. 
To date, more than 200 $z \gtrsim 6$ quasars have been discovered 
through various wide-field optical to near-infrared surveys, 
including SDSS \citep[e.g.,][]{2003AJ....125.1649F,2006AJ....131.1203F,2016ApJ...833..222J}, 
CFHQS \citep{2007AJ....134.2435W,2009AJ....137.3541W,2010AJ....139..906W}, 
VIKING \citep{2013ApJ...779...24V,2015MNRAS.453.2259V}, 
UKIDSS \citep{2009A&A...505...97M,2011Natur.474..616M}, 
Pan-STARRS1 \citep[e.g.,][]{2014AJ....148...14B,2016ApJS..227...11B,2017arXiv171001251M}, 
DES \citep[e.g.,][]{2017MNRAS.468.4702R}, 
DECaLS \citep{2017ApJ...839...27W}, 
SCam and HSC \citep{2015ApJ...798...28K,2016ApJ...828...26M,2017arXiv170405854M}, 
and several other projects \citep[e.g.,][]{2015MNRAS.451L..16C,2015Natur.518..512W}. 
Most of these surveys probed the bright end 
of the quasar population (UV magnitude $M_{\rm 1450} \lesssim -26$) 
powered by almost Eddington-limited mass accretion onto 
massive ($\gtrsim 10^9$ $M_\odot$) SMBHs \citep{2010AJ....140..546W,2014ApJ...790..145D}. 

It is very challenging to detect the rest-frame optical emission 
from the host galaxy of a quasar at high redshift ($z \gtrsim 4$), 
due to surface brightness dimming 
and the large brightness contrast \citep{2012ApJ...756L..38M,2012MNRAS.420.3621T}. 
However, cold gas and dust emission from star-forming regions 
have been used instead to probe the hosts 
at wavelengths relatively free from quasar emission. 
This approach has been advanced thanks to the advent of large and sensitive 
(sub)millimeter (hereafter sub/mm) interferometric arrays, 
such as the IRAM Plateau de Bure interferometer (now NOEMA) 
and the Atacama Large Millimeter/submillimeter Array (ALMA). 
Observations of galaxies hosting luminous $M_{\rm 1450} < -26$ quasars at $z \gtrsim 6$ 
have revealed large reservoirs of dust ($\sim 10^8~M_\odot$) 
and cold molecular gas ($\sim 10^{10}~M_\odot$) 
with high far-infrared (FIR) luminosities ($L_{\rm FIR} > 10^{12}~L_\odot$), 
indicating vigorous star formation activity 
(SFR $\gtrsim 100-1000~M_\odot$ yr$^{-1}$) coeval with the central AGNs 
\citep[e.g.,][]{2003A&A...406L..55B,2003A&A...409L..47B,2003AJ....126...15P,
2003MNRAS.344L..74P,2008MNRAS.383..289P,2004MNRAS.351L..29R,2004ApJ...615L..17W,
2007AJ....134..617W,2008ApJ...687..848W,2010ApJ...714..699W,2011AJ....142..101W,2011ApJ...739L..34W,
2013A&A...552A..43O,2014MNRAS.445.2848G,2015MNRAS.451.1713S}, 
and placing tight constraints on early star formation and dust formation histories 
\citep[e.g.,][]{2014MNRAS.438.2765C,2014MNRAS.444.2442V}. 

Interferometric studies of the strong 157.74 $\micron$ [C\,\emissiontype{II}] $^2$P$_{3/2}$ $\rightarrow$ $^2$P$_{1/2}$ emission line 
(rest frequency 1900.539 GHz), which is the principal coolant of 
photodissociation regions of galaxies \citep{1991ApJ...373..423S,1999RvMP...71..173H}, 
has also been an important tracer of the hosts of high-$z$ quasars. 
Such observations have revealed vigorous star-forming activity 
located in relatively compact regions (a few kpc in diameter) 
as well as the cold gas dynamics of $z \gtrsim 6$ galaxies, 
hosting not only luminous quasars \citep{2005A&A...440L..51M,2013ApJ...773...44W,
2016ApJ...830...53W,2015ApJ...805L...8B,2016ApJ...816...37V,2017arXiv171201886V,2017Natur.545..457D,2018arXiv180102641D,2017arXiv171001251M} 
but also less luminous ($M_{\rm 1450} \gtrsim -25$) quasars 
\citep{2013ApJ...770...13W,2015ApJ...801..123W,2017arXiv171002212W}. 
These dynamical studies revealed that $z \gtrsim 6$ luminous quasars 
have, on average, 10 times more massive SMBHs than 
the local co-evolutionary relations for a given velocity dispersion $\sigma$ or dynamical mass of the host 
\footnote{[C\,\emissiontype{II}] velocity dispersion ($\sigma_{\rm [C\,\emissiontype{II}]}$) is widely used 
as a surrogate for a stellar velocity dispersion in the case of high-$z$ quasars.}, 
implying that SMBHs were formed significantly earlier than their hosts \citep[e.g.,][]{2010MNRAS.405...29L}. 

However, there would be a selection bias for high redshift quasars 
toward more luminous objects or more massive SMBHs 
if the underlying $M_{\rm BH}$ distribution has a large scatter for a given galaxy mass 
\citep{2005ApJ...626..657W,2007ApJ...670..249L,2014MNRAS.438.3422S}. 
Therefore, it is vital to probe lower luminosity quasars and their hosts to obtain 
an unbiased view of early co-evolution that accounts for the bulk of the SMBH population 
at that time \citep[e.g.,][]{Schramm13}. 
Indeed, studies of less luminous ($M_{\rm 1450} \gtrsim -25$) CFHQS quasars 
with lower mass SMBHs ($\sim 10^8~M_\odot$) have revealed that 
their dynamical masses are well matched to those of local galaxies 
\citep{2015ApJ...801..123W,2017arXiv171002212W}. 
This lower luminosity regime is now being extensively explored 
with our wide-field and sensitive survey 
with the Hyper Suprime-Cam \citep[HSC;][]{2012SPIE.8446E..0ZM,Miyazaki17,Komiyama17,Kawanomoto17,Furusawa17} mounted on the Subaru telescope: 
we have discovered more than 50 quasars at $z \gtrsim 6$ \citep{2016ApJ...828...26M,2017arXiv170405854M}. 
We have organized an extensive multiwavelength follow-up consortium: 
Subaru High-z Exploration of Low-Luminosity Quasars (SHELLQs). 

In this paper, we report ALMA Cycle 4 observations of the 
[C\,\emissiontype{II}] 158 $\mu$m emission line 
and the underlying rest-frame FIR continuum emission 
of four HSC quasar host galaxies at $z \gtrsim 6$ (Table \ref{tbl1}), 
i.e., J0859+0022 (Ly$\alpha$-based redshift $z_{\rm Ly\alpha} = 6.39$), 
J1152+0055 ($z_{\rm Ly\alpha} = 6.37$), J2216-0016 ($z_{\rm Ly\alpha} = 6.10$), 
and J1202-0057 ($z_{\rm Ly\alpha} = 5.93$), originally discovered by \citet{2016ApJ...828...26M}. 
These HSC quasars are $\sim 3-4$ magnitudes fainter than 
most of the $z \gtrsim 6$ luminous-end quasars 
($M_{\rm 1450} < -26$) so far studied in the rest-FIR, 
and are comparably faint to the low-luminosity CFHQS quasars 
at $z \gtrsim 6$ \citep{2007AJ....134.2435W,2010AJ....139..906W}. 
We describe our observations in section 2. 
The observed properties of both [C\,\emissiontype{II}] line 
and FIR continuum emission are presented in section 3. 
Then we discuss the star-forming nature of the HSC quasar hosts 
and the less biased early co-evolution in section 4, 
and present our conclusions in section 5. 
Throughout the paper, we assume the standard cosmology 
with $H_0$ = 70 km s$^{-1}$ Mpc$^{-1}$, $\Omega_{\rm M}$ = 0.3, and $\Omega_{\rm \Lambda}$ = 0.7.

\section{Observations and data reduction}\label{sec2}
Four $z \gtrsim 6$ HSC quasars were observed during ALMA Cycle 4 
(ID = 2016.1.01423.S, PI: T. Izumi) at band 6 
between 2016 December 2 and 2017 April 13. 
Our observations are summarized in Table \ref{tbl1}, 
along with the basic target information. 
These observations were conducted in a single pointing 
(2 side-band dual-polarization mode) with $\sim 25\arcsec$ diameter field of view, 
which corresponds to $\sim 140$ kpc at the source redshifts 
(1$\arcsec$ corresponds to 5.5--5.8 kpc). 
The phase tracking centers were set to the optical quasar locations \citep{2016ApJ...828...26M}. 
The absolute positional uncertainty is $\sim 0''.1$ 
according to the ALMA Knowledgebase\footnote{https://help.almascience.org/index.php?/Knowledgebase/List}. 
With the minimum baseline length (15.1 m), 
the maximum recoverable scale of our observations is $\sim 9.5\arcsec$. 

The receivers were tuned to cover the redshifted [C\,\emissiontype{II}] line emissions 
whose frequencies were estimated from the measured redshifts of Ly$\alpha$. 
The total bandwidth of these observations was $\sim 7.5$ GHz, 
divided into four spectral windows of width 1.875 GHz. 
The native spectral resolution was 3.906 MHz (4.4--4.8 km s$^{-1}$), 
but 11--12 channels were binned to improve the signal-to-noise ratio ($S/N$), 
resulting in a final common velocity resolution of $\simeq 50$ km s$^{-1}$. 

Reduction and calibration of the data were performed 
with the Common Astronomy Software Applications (CASA) 
package \citep{2007ASPC..376..127M} version 4.7 in the standard manner. 
All images were reconstructed with the CASA task \verb|clean| 
(gain = 0.1, weighting = briggs, robust = 0.5). 
The achieved synthesized beams and rms sensitivities 
at a velocity resolution of 50 km s$^{-1}$ are summarized in Table \ref{tbl1}. 
All channels free of line emissions ($\sim 7.5$ GHz) were averaged 
to generate a continuum map for each source. 
The synthesized beams and rms sensitivities 
of these maps are also listed in Table \ref{tbl1}. 
For each source, the continuum emission was subtracted 
in the ($u$, $v$) plane before making the line cube. 
We used line intensities corrected for the primary beam attenuation for quantitative discussions, 
but this had a negligible effect, as all emission 
was found to lie in the central $r \lesssim 1.5\arcsec$ of each image. 
The pixel scale of all maps in this paper is set to 0$\arcsec$.1. 
Only statistical errors are displayed unless otherwise mentioned. 
Note that the systematic uncertainty of the absolute flux calibration at ALMA band 6 is 10\%, 
according to the ALMA Cycle 4 Proposer's Guide
\footnote{https://almascience.nao.ac.jp/proposing/documents-and-tools/cycle4/alma-proposers-guide}. 

\begin{longtable}{*{5}{c}}
\caption{Description of our sample and the ALMA observations}
\label{tbl1}
\hline\hline
 & J0859+0022 & J1152+0055 & J2216-0016 & J1202-0057 \\
\hline
\endhead
\hline
\endfoot
\hline
\multicolumn{5}{l}{{\bf Note.} Rest-frame UV properties are adapted from \citet{2016ApJ...828...26M,2017arXiv170405854M}.}\\
\endlastfoot
RA (J2000.0) & \timeform{08h59m07s.19} & \timeform{11h52m21s.27} & \timeform{22h16m44s.47} & \timeform{12h02m46s.37} \\ 
Dec (J2000.0) & $+$\timeform{00D22'55''.9} & $+$\timeform{00D55'36''.6} & $-$\timeform{00D16'50'.1} & $-$\timeform{00D57'01''.7} \\
$z_{\rm Ly\alpha}$ & 6.39 & 6.37 & 6.10 & 5.93 \\
$M_{\rm 1450}$ & $-$24.09 & $-$25.31 & $-$23.82 & $-$22.83 \\ \hline
Number of antennas & 45 & 45 & 40--41 & 42--45 \\ 
Baseline (m) & 15.1--704.1 & 15.1--704.1 & 15.1--704.1 & 15.1--492.0 \\ 
On-source time (minute) & 105 & 17 & 99 & 205 \\ 
Bandpass calibrator & J0854+2006 & J1229+0203 & J2148+0657 & J1229+0203 \\
Complex gain calibrator & J0909+0121 & J1220+0203 & J2226+0052 & J1220+0203 \\
Flux calibrator & J0750+1231, J0854+2006 & J1229+0203 & J2148+0657 & J1229+0203 \\ 
$T_{\rm sys}$ (K) & $\sim$80--100 & $\sim$70 & $\sim$80--110 & $\sim$90--150 \\ \hline
\multicolumn{5}{c}{[C\,\emissiontype{II}] cube}\\ \hline
Beam size: & 0$\arcsec$.64 $\times$ 0$\arcsec$.47 & 0$\arcsec$.52 $\times$ 0$\arcsec$.47 & 0$\arcsec$.54 $\times$ 0$\arcsec$.43 & 0$\arcsec$.79 $\times$ 0$\arcsec$.71 \\ 
Position Angle (East of North) & 60$\arcdeg$.9 & 72$\arcdeg$.3 & $-$62$\arcdeg$.2 & 79$\arcdeg$.4 \\
rms noise per 50 km s$^{-1}$ & \multirow{2}{*}{0.12} & \multirow{2}{*}{0.24} & \multirow{2}{*}{0.18} & \multirow{2}{*}{0.12} \\ 
(mJy beam$^{-1}$) &  &  &  &  \\ 
rms noise per 50 km s$^{-1}$ & \multirow{2}{*}{0.25} & \multirow{2}{*}{0.52} & \multirow{2}{*}{0.41} & \multirow{2}{*}{0.21} \\ 
(mJy beam$^{-1}$; 1$\arcsec$.5 aperture) &  &  &  &  \\ \hline
\multicolumn{5}{c}{Continuum map}\\ \hline
Observed continuum frequency (GHz) & 249.5 & 250.0 & 259.6 & 265.5 \\
Beam size: & 0$\arcsec$.56 $\times$ 0$\arcsec$.50 & 0$\arcsec$.54 $\times$ 0$\arcsec$.50 & 0$\arcsec$.56 $\times$ 0$\arcsec$.45 & 0$\arcsec$.81 $\times$ 0$\arcsec$.73 \\
Position Angle (East of North) & 61$\arcdeg$.2 & 70$\arcdeg$.1 & $-$62$\arcdeg$.2 & 80$\arcdeg$.0 \\ 
rms noise: & \multirow{2}{*}{9.5} & \multirow{2}{*}{20.7} & \multirow{2}{*}{13.2} & \multirow{2}{*}{8.8} \\ 
($\mu$Jy beam$^{-1}$) &  &  &  &  \\
rms noise: & \multirow{2}{*}{23.4} & \multirow{2}{*}{32.3} & \multirow{2}{*}{27.1} & \multirow{2}{*}{12.2} \\ 
($\mu$Jy beam$^{-1}$; 1$\arcsec$.5 aperture) &  &  &  &  \\
\end{longtable}

\section{Results}\label{sec3}
\begin{figure*}[hptb]
\begin{center}
\includegraphics[scale=0.475]{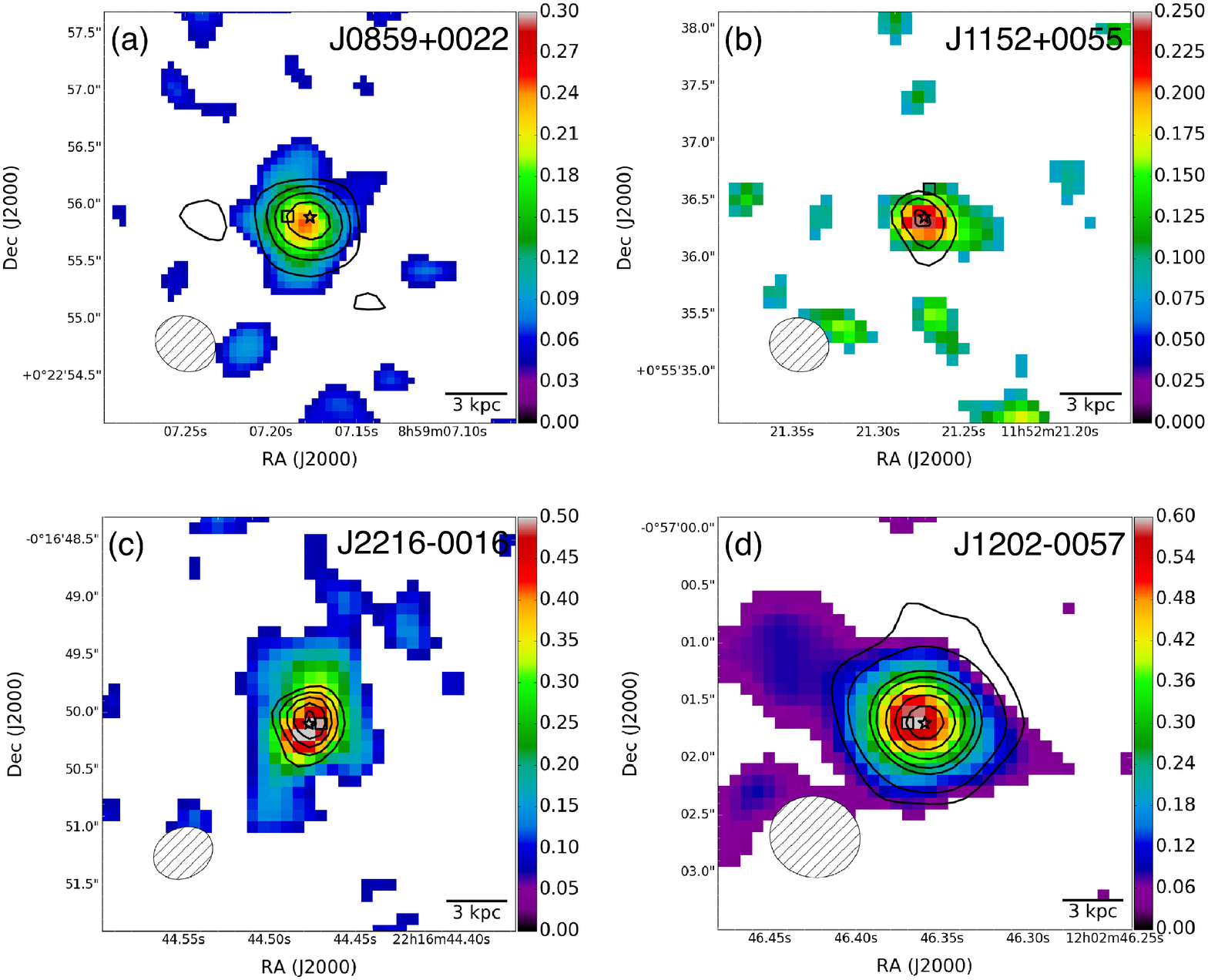}
\end{center}
\caption{
Spatial distributions of the velocity-integrated [C\,\emissiontype{II}] line (i.e., 0th moment map; color scale in Jy beam$^{-1}$ km s$^{-1}$ unit) 
and rest-frame FIR continuum (contours) emission of the HSC quasars, 
(a) J0859+0022, (b) J1152+0055, (c) J2216-0016, and (d) J1202-0057, 
visualized with the native resolutions. 
The synthesized beams are in the bottom-left corners. 
The central stars and the squares mark the continuum peaks 
at the rest-frame FIR (this work) and the rest-frame UV \citep{2016ApJ...828...26M}, respectively, 
which coincide within the positional uncertainties in every case. 
Contours indicate: (a) 3, 5, 7, 10$\sigma$ (1$\sigma$ = 9.5 $\mu$Jy beam$^{-1}$), 
(b) 3, 4, 5$\sigma$ (1$\sigma$ = 20.7 $\mu$Jy beam$^{-1}$), 
(c) 3, 4, 5, 6, 7$\sigma$ (1$\sigma$ = 13.2 $\mu$Jy beam$^{-1}$), and 
(d) 3, 5, 10, 12, 15, 18, 21$\sigma$ (1$\sigma$ = 8.8 $\mu$Jy beam$^{-1}$). 
The rms sensitivity of the velocity-integrated [C\,\emissiontype{II}] emission is, 
(a) 0.036, (b) 0.053, (c) 0.047, and (d) 0.027 Jy beam$^{-1}$ km s$^{-1}$, respectively. 
Pixels below these 1$\sigma$ levels were masked in the color maps. 
\label{fig1}} 
\end{figure*}

Figure \ref{fig1} shows the spatial distribution of 
the velocity-integrated (i.e., 0th moment) [C\,\emissiontype{II}] line 
and $\lambda_{\rm obs} \simeq 1.2$ mm (or $\lambda_{\rm rest} \simeq 158$ $\mu$m) 
continuum emission of the HSC quasars. 
Both line and continuum emission were clearly detected for all sources, 
with no apparent spatial offset among the line, 1.2 mm continuum, and optical centroids. 
Note that the velocity ranges which encompass the [C\,\emissiontype{II}] line emission 
were integrated over (using CASA  task \verb|immoments|) to make the moment-0 maps. 
The emission appears to be slightly extended relative to the synthesized beams. 
Given this, we decided to measure the rest-frame FIR properties 
with a common 1$\arcsec$.5 diameter circular aperture. 
The rms sensitivities within this aperture are also listed in Table \ref{tbl1}. 
The resultant properties are summarized in Table \ref{tbl2}. 

\begin{longtable}{*{5}{c}}
\caption{Rest-frame FIR properties of the HSC quasars \label{tbl2}}
\hline
\hline
 & J0859+0022 & J1152+0055 & J2216-0016 & J1202-0057 \\
\hline
\endhead
\hline 
\endfoot
\hline
\multicolumn{5}{l}{{\bf Note.} These were measured with a common 1$\arcsec$.5 aperture.}\\
\multicolumn{5}{l}{The (far-)infrared luminosities were estimated with a gray body spectrum model.}\\
\multicolumn{5}{l}{${\rm SFR}_{\rm [C\,\emissiontype{II}]}$/$M_\odot$~yr$^{-1}$ = 1.0 $\times$ 10$^{-7}$ ($L_{\rm [C\,\emissiontype{II}]}$/$L_\odot$)$^{0.98}$ \citep{2011MNRAS.416.2712D}.}\\
\multicolumn{5}{l}{${\rm SFR}_{\rm TIR}/M_\odot~{\rm yr^{-1}} = 1.49 \times 10^{-10} L_{\rm TIR}/L_\odot$ \citep{2011ApJ...737...67M}.}\\
\endlastfoot
$z_{\rm [C\,\emissiontype{II}]}$ & 6.3903 $\pm$ 0.0005 & 6.3637 $\pm$ 0.0005 & 6.0962 $\pm$ 0.0003 & 5.9289 $\pm$ 0.0002 \\
FWHM$_{\rm [C\,\emissiontype{II}]}$ (km s$^{-1}$) & 346 $\pm$ 46 & 192 $\pm$ 45 & 356 $\pm$ 33 & 335 $\pm$ 24 \\ 
$S_{\rm{[C\,\emissiontype{II}]}}$ (Jy km s$^{-1}$) & 0.45 $\pm$ 0.05 & 0.37 $\pm$ 0.08 & 1.05 $\pm$ 0.08 & 0.68 $\pm$ 0.04 \\ 
$L_{\rm{[C\,\emissiontype{II}]}}$ (10$^8$ $L_\odot$) & 4.6 $\pm$ 0.5 & 3.8 $\pm$ 0.8 & 10.2 $\pm$ 0.8 & 6.2 $\pm$ 0.4 \\ 
$f_{\rm 1.2mm}$ ($\mu$Jy) & 157 $\pm$ 23 & 189 $\pm$ 32 & 136 $\pm$ 27 & 246 $\pm$ 12 \\ 
EW$_{\rm{[C\,\emissiontype{II}]}}$ ($\mu$m) & 1.50 $\pm$ 0.28 & 1.02 $\pm$ 0.27 & 4.08 $\pm$ 0.87 & 1.44 $\pm$ 0.11 \\ 
SFR$_{\rm [C\,\emissiontype{II}]}$ ($M_\odot$ yr$^{-1}$) & 31 $\pm$ 4 & 25 $\pm$ 5 & 67 $\pm$ 5 & 42 $\pm$ 3 \\ \hline
\multicolumn{5}{c}{$T_d$ = 47 K, $\beta$ = 1.6}\\ \hline
$L_{\rm FIR}$ (10$^{11}$ $L_\odot$) & 3.4 $\pm$ 0.5 & 4.1 $\pm$ 0.7 & 2.8 $\pm$ 0.6 & 4.8 $\pm$ 0.2 \\
$L_{\rm TIR}$ (10$^{11}$ $L_\odot$) & 4.8 $\pm$ 0.7 & 5.8 $\pm$ 1.0 & 3.9 $\pm$ 0.8 & 6.7 $\pm$ 0.3 \\ 
SFR$_{\rm TIR}$ ($M_\odot$ yr$^{-1}$) & 71 $\pm$ 10  & 86 $\pm$ 14 & 58 $\pm$ 11 & 100 $\pm$ 5 \\
$M_{\rm dust}$ ($10^7~M_\odot$) & 2.4 $\pm$ 0.4 & 2.9 $\pm$ 0.5 & 2.0 $\pm$ 0.4 & 3.4 $\pm$ 0.2 \\
$L_{\rm [C\,\emissiontype{II}]}/L_{\rm FIR}$ ($10^{-3}$) & 1.4 $\pm$ 0.3 & 0.9 $\pm$ 0.2 & 3.7 $\pm$ 0.8 & 1.3 $\pm$ 0.1 \\ \hline
\multicolumn{5}{c}{$T_d$ = 35 K, $\beta$ = 1.6}\\ \hline
$L_{\rm FIR}$ (10$^{11}$ $L_\odot$) & 1.5 $\pm$ 0.2 & 1.7 $\pm$ 0.3 & 1.2 $\pm$ 0.2 & 2.0 $\pm$ 0.1 \\ 
$L_{\rm TIR}$ (10$^{11}$ $L_\odot$) & 1.9 $\pm$ 0.3 & 2.3 $\pm$ 0.4 & 1.5 $\pm$ 0.3 & 2.7 $\pm$ 0.1 \\ 
SFR$_{\rm TIR}$ ($M_\odot$ yr$^{-1}$) & 28 $\pm$ 4  & 34 $\pm$ 6 & 23 $\pm$ 5 & 40 $\pm$ 2 \\
$M_{\rm dust}$ ($10^7~M_\odot$) & 5.0 $\pm$ 0.7 & 6.0 $\pm$ 1.0 & 4.1 $\pm$ 0.8 & 7.1 $\pm$ 0.3 \\
$L_{\rm [C\,\emissiontype{II}]}/L_{\rm FIR}$ ($10^{-3}$) & 3.2 $\pm$ 0.6 & 2.2 $\pm$ 0.6 & 8.7 $\pm$ 1.8 & 3.1 $\pm$ 0.2 \\ 
\end{longtable}

\subsection{[C\,\emissiontype{II}] line properties}\label{sec3.1}
\begin{figure*}[hptb] 
\begin{center}
\includegraphics[scale=0.475]{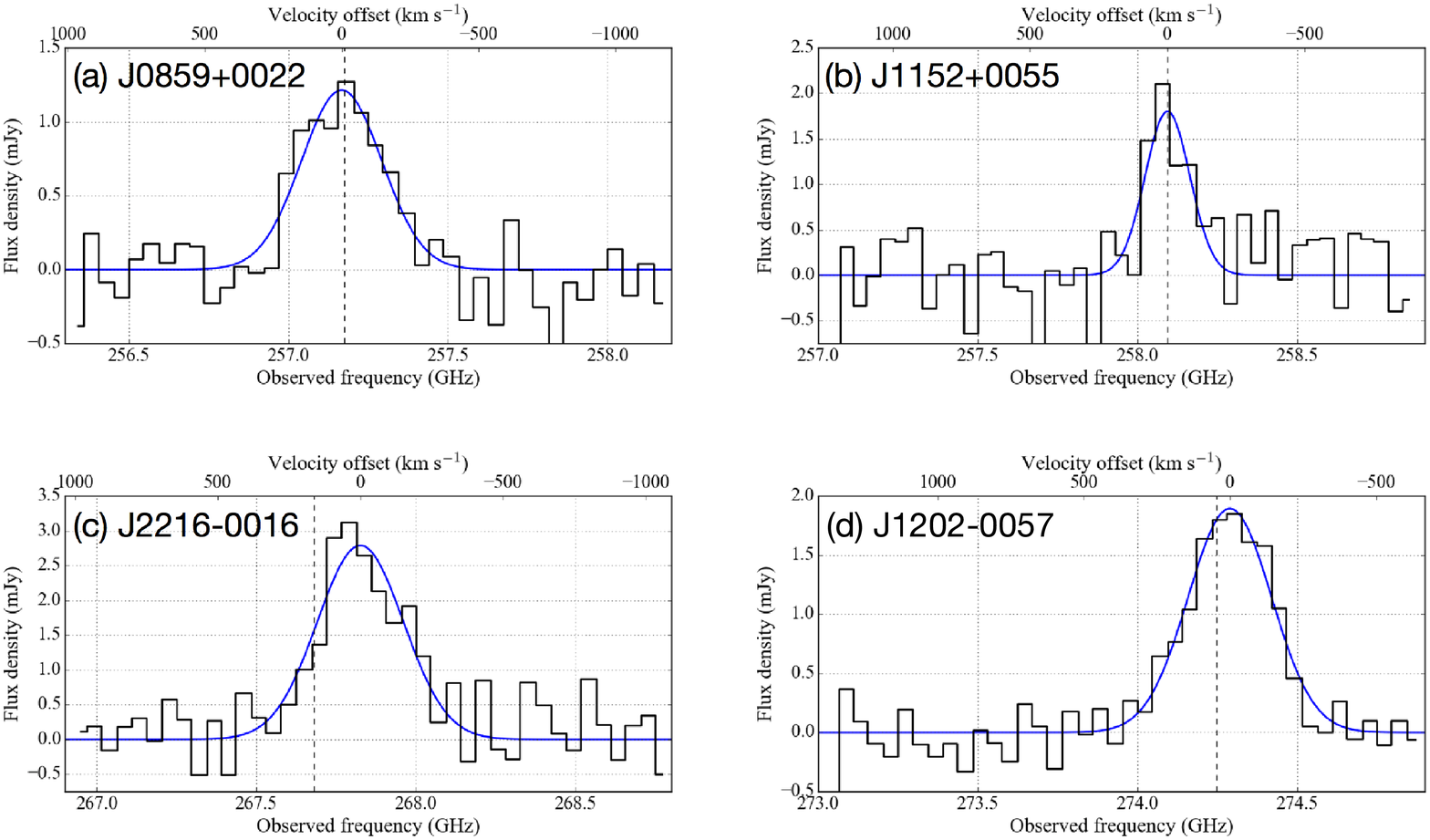}
\end{center}
\caption{
The [C\,\emissiontype{II}] line spectrum of the four HSC quasars obtained with ALMA. 
The blue curves indicate the best fit single Gaussian profiles. 
The upper axis in each panel is the velocity offset from the [C\,\emissiontype{II}] Gaussian peak. 
The expected [C\,\emissiontype{II}] frequencies from the Ly$\alpha$-based redshifts are indicated by the vertical dashed lines. 
}
\label{fig2}
\end{figure*}

Figure \ref{fig2} shows the [C\,\emissiontype{II}] line spectra 
measured with the 1$\arcsec$.5 diameter aperture. 
We fit each continuum-subtracted spectrum with a single Gaussian profile to extract 
the redshift ($z_{\rm [C\,\emissiontype{II}]}$), 
line width (full width at half maximum = FWHM$_{\rm [C\,\emissiontype{II}]}$), 
and the velocity-integrated line flux ($S_{\rm{[C\,\emissiontype{II}]}}$) 
of each source, as listed in Table 2. 
The line profiles were well fit with the single Gaussians 
(i.e., we found no strong indication of [C\,\emissiontype{II}] outflows), 
although we will examine the case of a double-Gaussian fit to J2216-0016 in subsection \ref{BAL}, 
as it is a broad absorption line quasar clearly showing nuclear outflows associated with it. 
However, as further observations are needed 
to confirm the necessity of the double-Gaussian fit, 
we use the results from the single Gaussian fit 
to be consistent across the sample. 

The velocity centroids of the [C\,\emissiontype{II}] lines show no significant offset 
from those determined based on their Ly$\alpha$ emission lines \citep{2016ApJ...828...26M}, 
even though the latter could have considerable uncertainties, 
due to intergalactic absorption \citep[e.g.,][]{2017ApJ...840...24E}. 
On the other hand, some quasars show $\gtrsim 500$ km s$^{-1}$ 
shifts (mostly blueshifts) of the Mg\,\emissiontype{II} line relative to the [C\,\emissiontype{II}] line 
\citep[e.g.,][]{2015ApJ...805L...8B,2015ApJ...801..123W,2016ApJ...816...37V,2016ApJ...830...53W,2017ApJ...836....8T}, 
suggesting fast ionized outflows at the nuclei of those quasars. 
The difference between $z_{\rm [C\,\emissiontype{II}]}$ and $z_{\rm Ly\alpha}$ 
is a measure of the neutral fraction of the intergalactic medium at $z > 6$, 
but this analysis is beyond the scope of this paper. 
The FWHMs of our HSC quasars are comparable to those 
of previously observed high-redshift quasars 
\citep[e.g.,][]{2013ApJ...773...44W,2015ApJ...801..123W,2016ApJ...816...37V}, 
which lie in the range of $\sim 200-500$ km s$^{-1}$. 

We also calculated the [C\,\emissiontype{II}] line luminosities of our sources 
with the standard equation of $L_{\rm [C\,\emissiontype{II}]} = 1.04 \times 10^{-3}~S_{\rm [C\,\emissiontype{II}]}~\nu_{\rm rest} ~(1+z_{\rm [C\,\emissiontype{II}]})^{-1} ~D^2_L$ \citep{2005ARA&A..43..677S}. 
Here, $L_{\rm [C\,\emissiontype{II}]}$ is the [C\,\emissiontype{II}] line luminosity in units of $L_\odot$, 
$\nu_{\rm rest}$ is in units of GHz, 
$S_{\rm [C\,\emissiontype{II}]}$ is in units of Jy~km~s$^{-1}$, 
and $D_L$ is the luminosity distance in units of Mpc, respectively. 
We obtained $L_{\rm [C\,\emissiontype{II}]} \simeq (4-10) \times 10^8$ $L_\odot$, 
with only the most luminous J2216-0016 reaching 10$^9$ $L_\odot$. 
This is in clear contrast to the corresponding values for optically luminous $z \gtrsim 6$ quasars, 
$L_{\rm [C\,\emissiontype{II}]} \simeq (1-10) \times 10^9~L_\odot$ 
\citep{2005A&A...440L..51M,2012ApJ...751L..25V,2016ApJ...816...37V,2017arXiv171201886V,
2013ApJ...773...44W,2016ApJ...830...53W,2015ApJ...805L...8B}. 
At these redshifts, FIR/submm-selected dusty starburst galaxies 
also exhibit $L_{\rm [C\,\emissiontype{II}]} \gtrsim (1-10) \times 10^{9}~L_\odot$ \citep{2013Natur.496..329R,2017ApJ...842L..15S}. 
On the other hand, the $L_{\rm [C\,\emissiontype{II}]}$ values for the HSC quasars are 
comparable to those of the less luminous CFHQS quasars 
\citep{2013ApJ...770...13W,2015ApJ...801..123W,2017arXiv171002212W}, 
as well as to those of UV/optically-selected galaxies at $z > 6$ \citep[e.g.,][]{2016ApJ...833...68A}. 

If we attribute the heating source of the [C\,\emissiontype{II}] 
emission to young stars, as is commonly assumed, 
the above trend indicates a lower SFR in the host galaxies of less luminous quasars. 
We then applied the relation in \citet{2011MNRAS.416.2712D} of 
${\rm SFR}_{\rm [C\,\emissiontype{II}]}$/$M_\odot$~yr$^{-1}$ = 1.0 $\times$ 
10$^{-7}$ ($L_{\rm [C\,\emissiontype{II}]}$/$L_\odot$)$^{0.98}$ 
to derive the values listed in Table 2. 
This relation, with a dispersion of $\sim 0.3$ dex (this is not included in the errors of the derived SFR), 
is calibrated with the Kroupa initial mass function \citep[IMF;][]{2001MNRAS.322..231K} 
for galaxies of $L_{\rm FIR} \lesssim 10^{12}~L_\odot$. 
It is consistent with that in \citet{2014ApJ...790...15S}, which is independently 
calibrated with infrared [Ne\,\emissiontype{II}] and [Ne\,\emissiontype{III}] lines \citep[see also][]{2015ApJ...800....1H}. 
The possible contribution of quasars to the [C\,\emissiontype{II}] line heating 
\citep[e.g.,][]{2010ApJ...724..957S} is neglected here, 
as (i) the so-called [C\,\emissiontype{II}]-deficit that would imply 
an influence of AGN on the FIR properties in cases of quasars was not seen in our sample (subsection 4.1), 
and (ii) the measured [C\,\emissiontype{II}] equivalent widths 
of the HSC quasars (EW$_{\rm [C\,\emissiontype{II}]}$; Table 2) 
are consistent with the typical values of local starburst galaxies 
\citep[e.g.,][]{2013ApJ...774...68D,2014ApJ...790...15S}. 
J2216-0016 even has an EW (4.08 $\pm$ 0.87 $\micron$) at the high-end of the range for local galaxies. 
Furthermore, a possible dependence of [C\,\emissiontype{II}] strength 
on the gas-phase metallicity \citep{2017arXiv171103735H} 
is also neglected as the HSC quasar hosts are found to be massive (section 4), 
implying that they would be evolved systems. 
The derived SFRs (25--67 $M_\odot$~yr$^{-1}$), and the $L_{\rm [C\,\emissiontype{II}]}$ values themselves, 
are well within the range of local luminous infrared galaxy (LIRG)-class systems, 
\citep[e.g.,][]{2011MNRAS.416.2712D,2013ApJ...774...68D,2014ApJ...790...15S}, 
although the HSC quasars are located at $z \gtrsim 6$.

\subsection{FIR continuum properties}\label{sec3.2}
The observed $\lambda = 1.2$ mm continuum emission ($f_{\rm 1.2mm}$; Table 2) 
is primarily emitted from the thermal dust 
\citep[e.g.,][]{2000ApJ...528..171Y,2001ApJ...555..625C,2006ApJ...642..694B}. 
We used this to derive the FIR \citep[42.5--122.5 $\mu$m;][]{1988ApJS...68..151H} continuum luminosity $L_{\rm FIR}$, 
which traces SFR if we assume that the cold dust is primarily heated 
by young stars \citep[e.g.,][]{1998ARA&A..36..189K}. 
This assumption is thought to be valid for quasars 
\citep[e.g.,][]{2006ApJ...649...79S,2014ApJ...785..154L}, 
although contrary arguments have also been proposed 
\citep[e.g.,][]{2016MNRAS.459..257S,2017MNRAS.465.1401S}. 
In practice, the intrinsic (i.e., AGN-heated) FIR spectral shape of quasars 
would vary from source to source \citep{2017ApJ...841...76L}, 
but handling of this effect is quite challenging at this moment. 

To compute $L_{\rm FIR}$, we first adopted an optically thin gray body spectrum model 
with dust temperature $T_d$ = 47 K and emissivity index $\beta$ = 1.6 (emissivity $\propto \nu^\beta$) 
to be consistent with previous $z > 6$ quasar studies 
\citep[e.g.,][]{2013ApJ...773...44W,2015ApJ...801..123W,2016ApJ...816...37V}. 
These fixed parameters are based on the mean spectral energy 
distribution of high-redshift optically/FIR-luminous quasars at $1.8 < z < 6.4$ 
\citep[][see also Leipski et al. 2014]{2006ApJ...642..694B}. 
However, it is uncertain whether these values 
are applicable to the much less luminous (at both the optical and FIR bands) HSC quasars, 
which should be studied further with future multiwavelengths observations. 
We also explore the consequence of lower $T_d$ below, which may be more realistic. 

We also considered the influence of the cosmic microwave background (CMB) 
on the submm observations at high redshifts \citep{2013ApJ...766...13D}, 
as the CMB provides additional source of dust heating. 
However, as those effects are negligible as long as we adopt $T_d \gtrsim 35$ K. 
We thus do not make any correction to the observed submm fluxes in this study: 
we should revise our estimation on, e.g., $L_{\rm FIR}$, once accurate $T_d$ is obtained. 

The resultant $L_{\rm FIR}$ ($T_d$ = 47 K) listed in Table 2. 
They all fall within a relatively narrow range, $\simeq (3-5) \times 10^{11}~L_\odot$, 
corresponding to the luminosity range of LIRGs. 
This is consistent with the [C\,\emissiontype{II}]-based results, 
where we also found LIRG-like line luminosities. 
The $L_{\rm FIR}$ of our HSC quasars are then 
much fainter than the $z \gtrsim 6$ optically luminous quasars by factors of $\simeq 10-100$ 
\citep[e.g.,][]{2007AJ....134..617W,2008ApJ...687..848W,2011AJ....142..101W}. 
On the other hand, their $L_{\rm FIR}$ are higher than some optically-selected 
normal galaxies (not AGN) at $z \sim 6$ that are not detected at FIR 
even with the high sensitivity of ALMA \citep{2015Natur.522..455C}, 
indicating that quasars are indeed dust-enriched systems. 

The SFR were simply estimated by 
(i) extending the gray body spectrum to the total-IR (TIR: 8--1000 $\mu$m, Table \ref{tbl2}) range, 
(ii) assuming that star-forming activity fully accounts for the TIR, 
and (iii) applying the conversion, ${\rm SFR}/M_\odot~{\rm yr^{-1}} = 1.49 \times 10^{-10} L_{\rm TIR}/L_\odot$ \citep{2011ApJ...737...67M}. 
This conversion is also grounded on the Kroupa IMF 
and is in accord with other studies \citep[e.g.,][]{1998ARA&A..36..189K} after accounting 
for the differing IMFs assumed therein. 
We found SFR = 58--100 $M_\odot$ yr$^{-1}$ for our sources. 
Note that we here neglected a contribution of UV luminosity to SFR estimate 
as a UV-to-IR luminosity ratio is small for massive star-forming 
galaxies such as having stellar mass of $\gtrsim 10^{10}~M_\odot$ \citep[e.g.,][]{2012ApJ...754L..29W,2017MNRAS.466..861D}, 
which would be the case for $z \gtrsim 6$ quasars (see also section 4). 
We also derived the cold dust mass ($M_{\rm dust}$), 
using $M_{\rm dust} = L_{\rm FIR}/(4\pi \int \kappa_\nu B_\nu d\nu)$ 
with a mass absorption coefficient $\kappa_\nu = \kappa_0 ~(\nu/250~{\rm GHz})^\beta$ 
and $\kappa_0 = 0.4$ cm$^2$ g$^{-1}$ \citep{2004A&A...425..109A,2006ApJ...642..694B}. 
Assuming $T_d = 47$ K, we found $M_{\rm dust} \simeq (2-3) \times 10^7~M_\odot$. 
As expected from our methodology, the values of SFR and $M_{\rm dust}$ are 
comparable to those observed in local LIRG-class systems 
\citep[e.g.,][]{2012ApJS..203....9U,2015ApJS..217....1T}. 

\begin{figure}[h] 
\begin{center}
\includegraphics[scale=0.2]{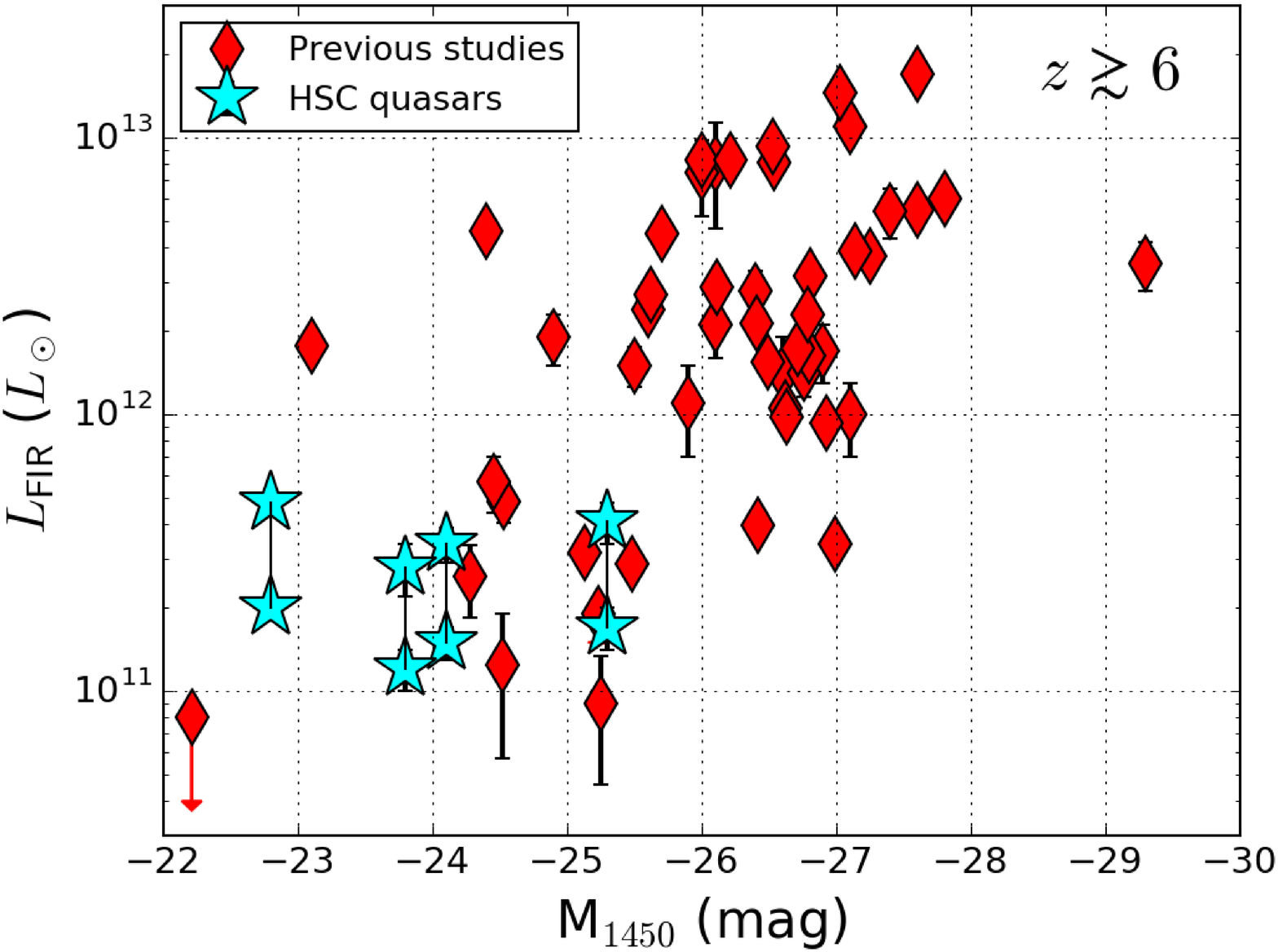}
\end{center}
\caption{
FIR luminosity ($L_{\rm FIR}$) as a function of quasar UV absolute magnitude ($M_{\rm 1450}$) 
for our HSC quasars (cyan stars) along with previously studied $z \gtrsim 6$ quasars 
\citep{2005A&A...440L..51M,2012ApJ...751L..25V,2016ApJ...816...37V,2017arXiv171201886V,2013ApJ...773...44W,2016ApJ...830...53W,2013ApJ...770...13W,2015ApJ...801..123W,2017arXiv171002212W,2015ApJ...805L...8B,2017arXiv171201860B,2017Natur.545..457D,2018arXiv180102641D,2017arXiv171001251M}. 
$L_{\rm FIR}$ is calculated with the gray body model with $T_d$ = 47 K and $\beta$ = 1.6. 
For the HSC quasars, we also plot the cases with $T_d$ = 35 K. 
This figure enhances the fact that we are probing lower-luminosity population 
in terms of both optical and FIR luminosity at $z \gtrsim 6$ with the HSC survey. 
}
\label{fig_add}
\end{figure}

However, the above TIR-based SFR (SFR$_{\rm TIR}$) 
are systematically larger than the [C\,\emissiontype{II}]-based 
SFR (SFR$_{\rm [C\,\emissiontype{II}]}$) except for the case of J2216-0016. 
This could not be due to significant AGN contamination to the dust heating, 
as the observed EW$_{\rm [C\,\emissiontype{II}]}$ values are comparable to star-forming LIRGs. 
On the other hand, $T_d$ itself is usually very uncertain even among star-forming galaxies: 
our less luminous HSC quasars (at both the optical and FIR bands) 
may have lower $T_d$ than the luminous-end quasars. 
If we adopt $T_d = 35$ K instead, which is a typical value 
observed in local LIRGs \citep[e.g.,][]{2012ApJS..203....9U} 
and SMGs at $z \sim 1-3$ having U/LIRG-class $L_{\rm FIR}$ 
\citep[e.g.,][]{2005ApJ...622..772C,2006ApJ...650..592K,2008MNRAS.384.1597C}, 
the resultant $L_{\rm FIR}$ is reduced by $\simeq 50-60$\%, by fixing $\beta$ to 1.6. 
In this case (Table \ref{tbl2}), the relevant FIR continuum properties are then, 
$L_{\rm FIR} \simeq (1-2) \times 10^{11} L_\odot$, 
SFR$_{\rm TIR}$ = 23--40 $M_\odot$ yr$^{-1}$, 
$M_{\rm dust} \simeq (4-7) \times 10^7 M_\odot$, respectively. 
The SFR$_{\rm TIR}$ now agrees better with SFR$_{\rm [C\,\emissiontype{II}]}$ 
for J0859+0022, J1152+0055, and J1202-0057, 
although the discrepancy in $L_{\rm FIR}$ between the HSC quasars 
and the optically-luminous quasars becomes larger (Figure \ref{fig_add}). 

\subsection{Spatial extent of the star-forming region}\label{sec3.3} 
The spatial extent of the star-forming region
\footnote{This could have a different distribution from the already existing stellar component 
\citep[see recent discussion in, e.g.,][]{2015ApJ...807..128S}.} 
of each source was estimated by applying 
a two-dimensional Gaussian fit to both the [C\,\emissiontype{II}] 
integrated intensity and the continuum maps, using the CASA task \verb|IMFIT|. 
This method (i.e., image-plane fitting) is consistent with those 
used in the previous submm studies on quasar host galaxies, 
which enables us a fair comparison of our results with them. 
We note another possibility to estimate the source sizes by using $uv$-plane fitting, 
but such a method only applies for data with decent $S/N$ 
\citep[$\sim 10-15$, e.g.,][]{2015ApJ...810..133I}, 
which is not the case for our work. 

The native resolution data (Table \ref{tbl1} and Figure \ref{fig1}) were used for our size measurements 
with 3$\sigma$ clipping to avoid noise contamination. 
The resultant values are listed in Table \ref{tbl3}: 
the [C\,\emissiontype{II}] emitting regions have sizes of FWHM $\sim$2.6--5.2 kpc, 
while the associated uncertainties are still large. 

We found good consistency between the [C\,\emissiontype{II}]-based sizes 
and continuum-based sizes within the uncertainties, except for the case of J2216-0016, 
where the [C\,\emissiontype{II}]-based size is $\sim 1.8$ times larger 
than the continuum-based size along its major axis (see also Figure \ref{fig1} and subsection 3.5). 
Interestingly, the spatial extents of our HSC quasars (even including J2216-0016) 
are comparable to those of the $z \gtrsim 6$ optically-luminous quasars observed at submm, 
having $M_{\rm dust}$ $\sim$ several $\times$ 10$^8~M_\odot$ \citep[e.g.,][]{2013ApJ...773...44W,2016ApJ...816...37V}, within the uncertainties. 
Thus, an order of magnitude difference in SFR (or $M_{\rm dust}$) between 
these populations could directly translate to a similar 
level of difference in the SFR \citep[or ISM mass 
under a certain gas-to-dust mass ratio, e.g.,][]{2007ApJ...663..866D} {\it surface density}. 
Indeed, gas mass surface density around an AGN is a key parameter for the black hole mass accretion, 
as it controls the gravitational instability therein, 
which can transport gas inward \citep[e.g.,][]{2010MNRAS.407.1529H}.

\begin{longtable}{*{3}{c}}
\caption{Spatial extent of the star-forming region of the HSC quasars}\label{tbl3}
\hline
\hline
Name & Size ([C\,\emissiontype{II}] FWHM) & Size (continuum FWHM) \\ \hline
\endhead
\hline 
\endfoot
\hline
\endlastfoot
  \multirow{2}{*}{J0859+0022} & (0$\arcsec$.51 $\pm$ 0$\arcsec$.15) $\times$ (0$\arcsec$.33 $\pm$ 0$\arcsec$.19) & (0$\arcsec$.39 $\pm$ 0$\arcsec$.16) $\times$ (0$\arcsec$.27 $\pm$ 0$\arcsec$.15) \\ 
   & (2.8 $\pm$ 0.8) kpc $\times$ (1.8 $\pm$ 1.0) kpc & (2.2 $\pm$ 0.9) kpc $\times$ (1.5 $\pm$ 0.8) kpc \\ \hline   
  \multirow{2}{*}{J1152+0055} & (0$\arcsec$.58 $\pm$ 0$\arcsec$.20) $\times$ (0$\arcsec$.25 $\pm$ 0$\arcsec$.13) & (0$\arcsec$.61 $\pm$ 0$\arcsec$.19) $\times$ (0$\arcsec$.24 $\pm$ 0$\arcsec$.17) \\
   & (3.0 $\pm$ 1.1) kpc $\times$ (1.4 $\pm$ 0.7) kpc & (3.4 $\pm$ 1.0) kpc $\times$ (1.3 $\pm$ 0.9) kpc \\ \hline
  \multirow{2}{*}{J2216-0016} & (0$\arcsec$.91 $\pm$ 0$\arcsec$.15) $\times$ (0$\arcsec$.44 $\pm$ 0$\arcsec$.12) & (0$\arcsec$.52 $\pm$ 0$\arcsec$.16) $\times$ (0$\arcsec$.41 $\pm$ 0$\arcsec$.23) \\
   & (5.2 $\pm$ 0.8) kpc $\times$ (2.5 $\pm$ 0.7) kpc & (2.9 $\pm$ 0.9) kpc $\times$ (2.3 $\pm$ 1.3) kpc \\ \hline
  \multirow{2}{*}{J1202-0057} & (0$\arcsec$.45 $\pm$ 0$\arcsec$.12) $\times$ (0$\arcsec$.27 $\pm$ 0$\arcsec$.19) & (0$\arcsec$.47 $\pm$ 0$\arcsec$.10) $\times$ (0$\arcsec$.42 $\pm$ 0$\arcsec$.14) \\ 
   & (2.6 $\pm$ 0.7) kpc $\times$ (1.5 $\pm$ 1.1) kpc & (2.7 $\pm$ 0.6) kpc $\times$ (2.4 $\pm$ 0.8) kpc \\ \hline
\end{longtable}

\subsection{Other emitters within the fields}\label{sec3.4} 
We searched for other emitters within the fields of view (FoV; HPBW $\simeq 25\arcsec$) of each source. 
The line cubes and the continuum maps with the native angular resolutions were used here. 

\subsubsection{[C\,\emissiontype{II}] emitters adjacent to the quasars?}
We first searched for [C\,\emissiontype{II}] emitters within the FoVs, 
particularly those associated with the central quasars. 
We applied the procedure described in \citet{2017ApJ...845..108Y} 
to our continuum-subtracted cubes. 
To this end, spectral windows (1.875 GHz width with $\simeq 43$ MHz binning) 
that contain the quasar [C\,\emissiontype{II}] emission were surveyed. 
We used the \verb|CLUMPFIND| software \citep{1994ApJ...428..693W} 
to search for line emitters other than the HSC quasars themselves with a peak $S/N \geq 5$. 
The relevant parameters were $\Delta~S = 1\sigma$ and $S_{\rm start} = 3\sigma$, 
where $\Delta~S$ is the contouring interval, 
and $S_{\rm start}$ is the starting contour level. 
To avoid spurious detections, we rejected candidates 
that do not show $S/N \geq 3$ emission in any of the channels adjacent to the peak ones. 
The frequency resolution (43 MHz) or the velocity resolution 
of 50 km s$^{-1}$ for objects near the quasar redshifts would be sufficient to 
resolve a typical line width \citep[$\sim 150-200$ km s$^{-1}$; e.g.,][]{2016ApJ...833...68A} 
of galaxies at $z \sim 6$ into several spectral elements. 

As a result, we did not detect any significant line emitter within the FoVs. 
This result holds if we decrease the detection threshold to 4.5$\sigma$, 
which is roughly consistent with the negative tail of the $S/N$ distributions of our cubes. 
Note that \citet{2016ApJ...816...37V} also reported a non-detection of other line emitters 
within the ALMA band 6 FoVs of three luminous $z > 6$ VIKING quasars. 
Regarding luminous [C\,\emissiontype{II}] emitters ($L_{\rm [C\,\emissiontype{II}]} \gtrsim 10^9~L_\odot$), 
our non-detection is broadly consistent with \citet{2017Natur.545..457D} as well, 
who reported four companion luminous [C\,\emissiontype{II}] emitting galaxies 
out of 25 luminous quasar fields at $z \gtrsim 6$: 
based on their detection rate (16\%), the expected number of 
such luminous [C\,\emissiontype{II}] emitters for our observations is at most one. 
Furthermore, \citet{2017ApJ...836....8T} reported three detections 
of associated luminous SMGs ($L_{\rm [C\,\emissiontype{II}]} \gtrsim 10^9~L_\odot$) 
out of six luminous quasar fields at $z \simeq 4.8$, 
implying a higher merger frequency at that redshift than at $z \sim 6$. 
As compared to those observations \citep{2017Natur.545..457D,2017ApJ...836....8T}, 
our observations are much deeper. 
Therefore, the non-detection of [C\,\emissiontype{II}] emitters in our fields 
would place a more stringent constraint on a [C\,\emissiontype{II}] luminosity function at that redshift, 
which is although beyond the scope of this paper.

\subsubsection{Continuum emitters}
\begin{figure}[h] 
\begin{center}
\includegraphics[scale=0.4]{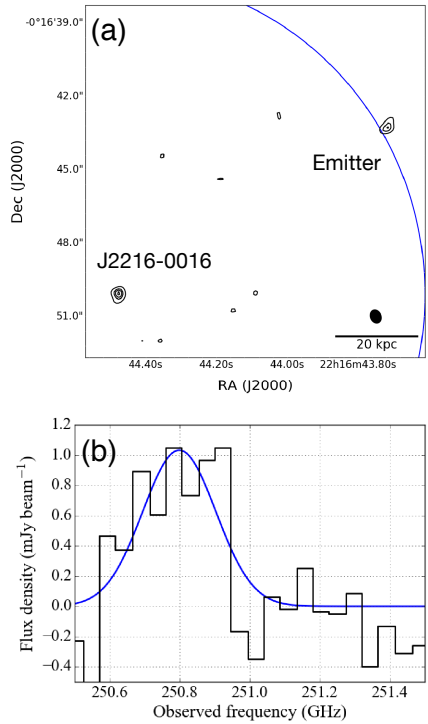}
\end{center}
\caption{
(a) A line emitter found in the J2216-0016 field that is 13$\arcsec$.2 away from the quasar. 
The blue line indicates the nominal field of view ($\sim 12\arcsec$ radius) of this ALMA band 6 observation. 
The contours step as 3, 5, 6, and 7$\sigma$ (1$\sigma$ = 13.2 $\mu$Jy beam$^{-1}$). 
The synthesized beam is plotted in the bottom-right corner. 
(b) The observed spectrum of the line emitter shown in (a), 
with the single Gaussian fit superposed. See text for the result of the fit. 
}
\label{fig3}
\end{figure}

With the rms values listed in Table \ref{tbl1}, we conservatively considered 
sources with a signal-to-noise ratio ($S/N$) $\geq 5$ as continuum emitters. 
In the fields of J0859+0022 (5$\sigma$ = 48$\mu$Jy beam$^{-1}$), 
J1152+0055 (104 $\mu$Jy beam$^{-1}$), and J1202-0057 (44 $\mu$Jy beam$^{-1}$), 
no significant emitter was found. 
In contrast, one emitter {\it candidate} was found 
slightly outside the nominal FoV of J2216-0016 (5$\sigma$ = 66 $\mu$Jy beam$^{-1}$; Figure \ref{fig3}a), 
13$\arcsec$.2 away from the quasar. 
Its coordinates are R.A. = \timeform{22h16m43s.718}, Dec. = $-$\timeform{00D16'43''.28}. 

However, this source turned out to be a line emitter, 
after a careful inspection of all spectral windows (Figure \ref{fig3}b), rather than a continuum emitter. 
That is, this source is not detected in the continuum map of J2216-0016 
after the channels with faint line emission were removed. 
A single Gaussian fit to the line gave an amplitude, centroid, and integrated intensity of 
1.03 $\pm$ 0.09 mJy, 250.80 $\pm$ 0.01 GHz, and 0.32 $\pm$ 0.03 Jy km s$^{-1}$, respectively. 
We suggest that the source is a lower-redshift object, 
as it is detected in all of the HSC bands as 
$g = 26.90 \pm 0.26$ mag, $r = 25.59 \pm 0.18$ mag, 
$i = 24.59 \pm 0.06$ mag, $z = 23.82 \pm 0.09$ mag, and $y = 23.10 \pm 0.06$ mag, 
which is typically not the case for high-redshift galaxies. 
Indeed, the HSC photometric-redshift catalog from the first data release 
\citep[with the Mizuki-code,][]{2015ApJ...801...20T,2017arXiv170405988T} 
suggests that the source is at $z_{\rm photo} = 1.32 \pm 0.11$. 
In this case, the line could be CO(5--4) emission (expected redshift $z_{\rm CO(5-4)} = 1.298 \pm 0.003$). 

In short, we did not find any significant continuum emitter 
in these four fields (each has $\simeq 0.135$ arcmin$^2$ field of view; $\sim 0.54$ arcmin$^2$ total) 
even at our high sensitivities (5$\sigma$ = 44--104 $\mu$Jy beam$^{-1}$). 
It is noteworthy, on the other hand, that many studies 
suggest that luminous quasars tend to reside 
in over-dense region of continuum emitters 
\citep[e.g.,][and references therein]{2015ApJ...806L..25S}: 
our result of the lower luminosity quasars at $z \gtrsim 6$ 
seems not to match those findings. 
Further studies on, e.g., halo masses of those less-luminous quasars, 
are needed to reveal underlying physical differences between 
the environments of various kinds of quasars. 

The non-detection in the four fields seems to be lower 
than the expectation from recent 1.2 mm number counts 
\citep[e.g.,][]{2016ApJ...833...68A,2016ApJS..222....1F}. 
For example, the cumulative number count in \citet{2016ApJS..222....1F} 
is $\sim 8$ (total number) in our four fields ($\sim 2-3$ in each 0.135 arcmin$^2$ field), 
whilst no source was detected. 
However, the discrepancy in each field is still statistically 
not so significant \citep[$\sim 1-2\sigma$;][]{1986ApJ...303..336G}: 
we would also suggest several possible factors that can further reconcile the discrepancy. 
One factor is cosmic variance given the small area we probed. 
A low selection-completeness expected for high resolution 
(e.g., $\sim 0\arcsec.5$) observations \citep[see Figure 5 in][]{2017ApJ...850...83F} will reduce the detection rate. 
It is also possible that previous counts using low-significance detections 
are contaminated by noise components, 
as suggested by \citet{2016ApJ...822...36O} and \citet{2017ApJ...835...98U}. 
{\it Pseudo continuum emitters}, which we found in the J2216-0016 field (Figure \ref{fig3}), 
could also be contaminants, particularly at the faint end. 
The accumulation of datasets of individually very deep observations 
will help to further constrain the true mm/submm 
number count at the faint end ($\lesssim 100$ $\mu$Jy): 
this will be investigated with our growing SHELLQs sample. 

\subsection{Any peculiarity in the BAL quasar J2216-0016?}\label{BAL}
Among the four HSC quasars studied here, J2216-0016 
shows a clear broad absorption line (BAL) feature \citep[e.g.,][]{1991ApJ...373...23W,2006ApJS..165....1T}
in its N\,\emissiontype{V} spectrum \citep{2016ApJ...828...26M}. 
As the BAL feature is a clear manifestation of nuclear outflows, 
which could be an indication of AGN-feedback on the host galaxy, 
it is interesting to see if there is any peculiarity in the host galaxy. 
Indeed, the relatively large spatial extent of the [C\,\emissiontype{II}] emitting region (5.2 $\pm$ 0.8 kpc; Table \ref{tbl3}) 
as well as the high EW$_{\rm [C\,\emissiontype{II}]}$ \citep[4.08 $\pm$ 0.87 $\micron$, 
compared to the value of $\simeq 1.0$ $\micron$ for local starburst galaxies;][]{2014ApJ...790...15S} 
already stands out among the four HSC quasars. 

\begin{figure}[h] 
\begin{center}
\includegraphics[scale=0.26]{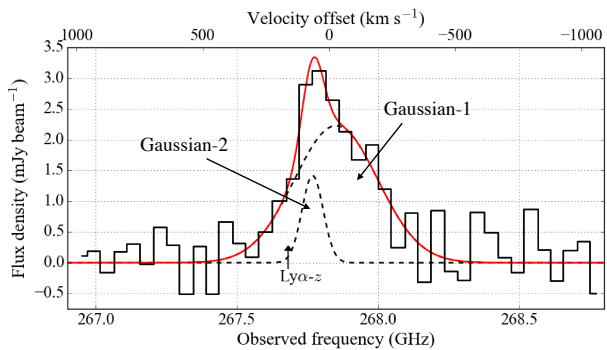}
\end{center}
\caption{
A double-Gaussian fit to the [C\,\emissiontype{II}] spectrum of J2216-0016 (red-solid line) 
with the black-dashed lines indicating each component. 
The upper axis denotes the velocity offset from the centroid of the single Gaussian profile shown in Figure \ref{fig2}. 
The expected [C\,\emissiontype{II}] frequency from the Ly$\alpha$-based redshift is also indicated. 
}
\label{fig4}
\end{figure}

\begin{figure}[h] 
\begin{center}
\includegraphics[scale=0.3]{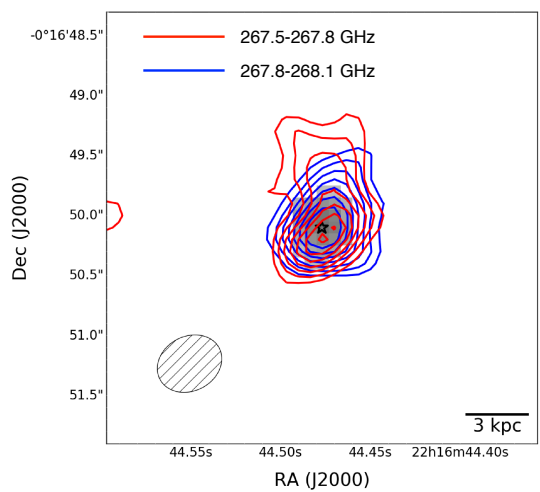}
\end{center}
\caption{
The integrated [C\,\emissiontype{II}] line emission (moment 0) map of J2216-0016 
divided into the redder half (red contours, containing the Gaussian-2; 267.5--267.8 GHz) 
and the bluer half (blue contours, containing the Gaussian-1; 267.8--268.1 GHz), 
overlaid on the continuum map (grayscale; Figure \ref{fig1}). 
The contours step as 3, 4, 5, ..., 8, and 9$\sigma$ with 1$\sigma$ = 0.023 Jy beam$^{-1}$ km s$^{-1}$. 
There is no significant offset between the peaks. 
The central star marks the peak location of the rest-FIR continuum emission. 
The synthesized beam is shown in the bottom-left corner. 
}
\label{fig5}
\end{figure}

As an initial investigation, we fit the observed [C\,\emissiontype{II}] spectrum 
with a double-Gaussian profile (Figure \ref{fig4}): 
the resultant reduced $\chi^2$ is 1.05 (degree of freedom
\footnote{We evaluated the fit within the frequency range of 267.41--268.30 GHz.}, d.o.f. = 14), 
which is improved over the single-Gaussian fit (reduced $\chi^2$ = 1.24 with d.o.f = 17). 
The two double-Gaussian constituents have centroid frequencies of 
267.851 GHz (Gaussian-1) and 267.768 GHz (Gaussian-2), respectively. 
The Gaussian-2 has a much narrower FWHM (99.1 km s$^{-1}$) than the Gaussian-1 (389.7 km s$^{-1}$) has. 
It is therefore plausible that the Gaussian-1 component corresponds to the quasar host, 
once we take into account the [C\,\emissiontype{II}] FWHM of the quasar host galaxies studied so far. 
Figure \ref{fig5} shows the integrated intensity of the [C\,\emissiontype{II}] emission 
divided into the redder and bluer velocity components. 
The redder component that contains the Gaussian-2 
is spatially more extended than the bluer one, toward the north to northeast direction. 
That elongation has a different position angle (P.A.) of $\sim 15^\circ$ 
from the bluer component, and most of the redder component 
around the continuum peak (P.A. $\sim -25^\circ$ to $-40^\circ$), 
implying that the elongation traces 
a physically different structure from the quasar host galaxy. 

One plausible origin for the elongated structure is an AGN-driven [C\,\emissiontype{II}] outflow, 
such as that observed in the luminous $z = 6.42$ quasar SDSS J1148+5251 
\citep{2012MNRAS.425L..66M,2015A&A...574A..14C}. 
In this case, the decomposed line profile suggests that the outflow 
is single-sided relative to the systemic velocity of the quasar host, 
although its width is rather narrow as compared to 
previously observed [C\,\emissiontype{II}] outflows. 
Similar single-sided outflows have been observed in many systems 
\citep[e.g.,][]{2014A&A...562A..21C,2017arXiv170605527F}. 
Note that the blueshifted BAL absorption feature 
does not necessarily preclude the existence of redshifted outflows. 

Another plausible origin of the offset is a galaxy merging with the quasar host 
with a projected separation of $\lesssim 5$ kpc. 
High resolution and deep optical/infrared imaging data, 
which is not available at this moment, 
will give a crucial hint on the nature of the extended component. 
A similarly close (projected distance $\sim 5$ kpc) galaxy was also found 
near the moderate luminosity ($M_{\rm 1450} = -25.6$) 
quasar PSO J167-13 \citep{2017arXiv171002212W}, 
which demonstrates the importance of high angular resolution in unveiling such close companion(s). 
If we suppose that the merging system has the same spatial extent as the quasar host itself, 
the much narrower FWHM roughly translates to dynamical mass an order of magnitude smaller, 
i.e., this event will be a minor merger, which would enhance nuclear activity 
\citep[e.g.,][]{1999ApJ...524...65T,2014MNRAS.440.2944K}. 
Such an evolutionary link between the BAL and early galaxy evolution 
is an appealing topic for further investigations \citep[e.g.,][]{2007ApJ...662L..59F}. 

In either case, we clearly need higher resolution and sensitivity 
to spatially isolate the candidate structure and to distinguish these two scenarios, 
particularly via studying the gas dynamics. 
In what follows, we use dynamical properties from the single Gaussian fit.

\section{Discussion}\label{sec4}
In this section, we first explore the star-forming nature of the HSC quasar host galaxies, 
and then discuss the early co-evolutionary relationship at $z \sim 6$, 
paying attention to the physical differences between optically luminous ($M_{\rm 1450} \lesssim -26$) 
and low-luminosity ($M_{\rm 1450} \gtrsim -25$) quasars. 
We will also compare the observed properties with theoretical predictions 
from semi-analytic models, particularly those from a {\it new numerical galaxy catalog} 
\citep[= $\nu^2$GC;][]{2016PASJ...68...25M}. 
In this catalog, the underlying merging histories of dark matter haloes 
are based on state-of-the-art cosmological $N$-body simulations \citep{2015PASJ...67...61I}, 
which have high mass resolution and quite large volumes relative to previous simulations, 
which are particularly suitable to study statistical properties 
of rare populations such as massive/luminous quasars at high redshifts (H. Shirakata et al. in preparation). 
The $\nu^2$GC simulation uses prescriptions for star formation, gas heating by UV feedback, 
supernova feedback, SMBH growth, and AGN feedback 
\citep[see also][]{2014ApJ...794...69E,2015MNRAS.450L...6S}, to trace galaxy evolution. 
In this study, we adopt the results from a subset of the $\nu^2$GC with the largest volume, 
i.e., $\nu^2$GC-L \citep{2015PASJ...67...61I}, 
for which the box size is 1.12 $h^{-1}$ cGpc and the dark-matter 
mass resolution is 2.20 $\times$ 10$^8$ $h^{-1}~M_\odot$ 
(number of particles = 8192$^3$). 

\subsection{The [C\,\emissiontype{II}]-FIR luminosity relation}
\begin{figure}[h] 
\begin{center}
\includegraphics[scale=0.215]{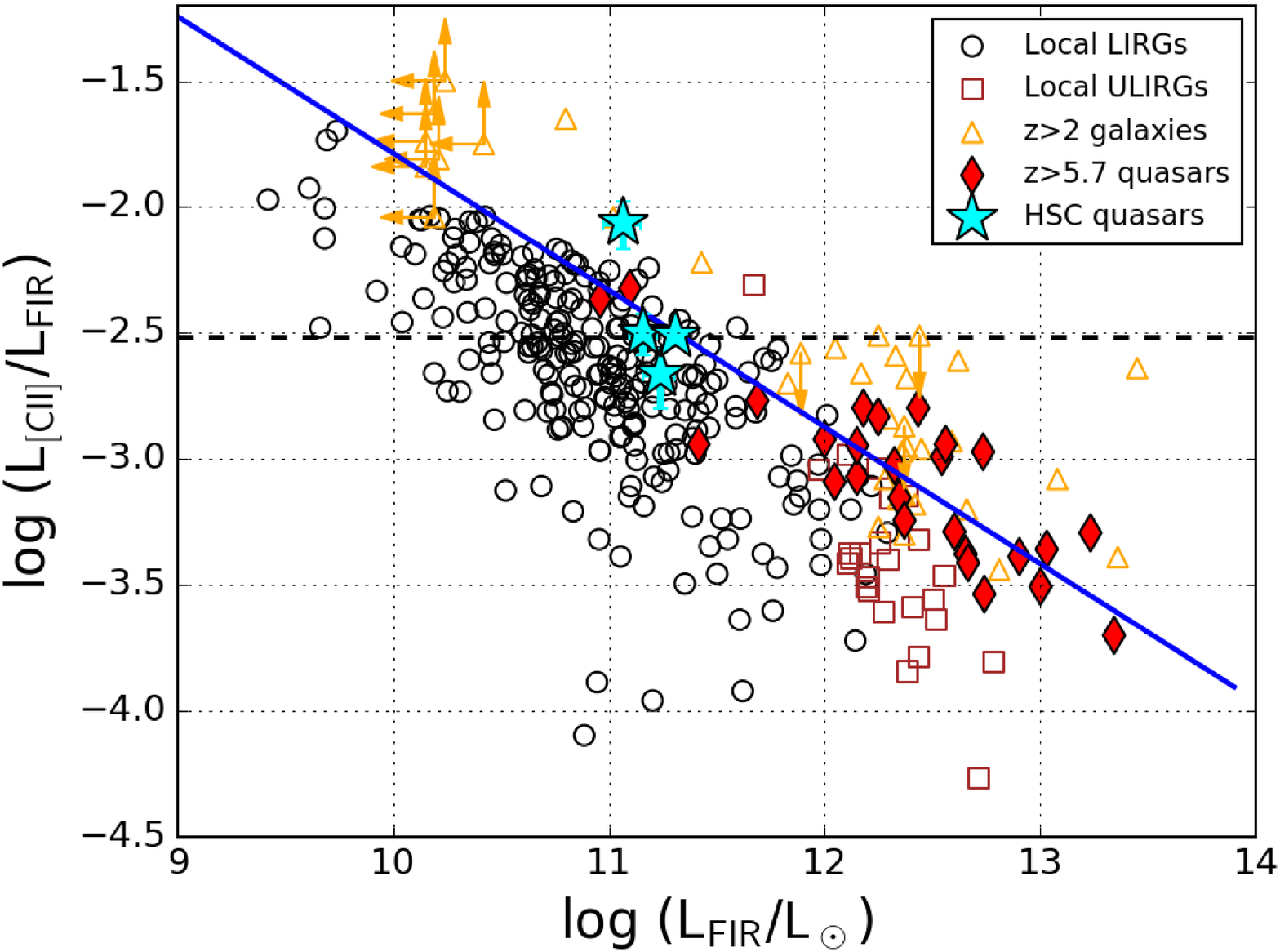}
\end{center}
\caption{
[C\,\emissiontype{II}]/FIR luminosity ratio as a function of FIR luminosity for our HSC quasars (cyan stars). 
Also plotted are compilations of various kinds of galaxies from the recent literature for a comparison: 
local LIRGs \citep{2013ApJ...774...68D}, 
local ULIRGs \citep{2013ApJ...776...38F}, 
$z > 2$ star-forming galaxies \citep[mostly FIR-selected SMGs, plus 
some UV-selected galaxies,][]{2009A&A...500L...1M,2010A&A...518L..35I,
2011A&A...530L...8D,2012ApJ...752L..30W,2013Natur.496..329R,2015MNRAS.449.2883G,2015Natur.522..455C}, 
and $z > 5.7$ quasars \citep{2005A&A...440L..51M,2013ApJ...773...44W,2016ApJ...830...53W,
2013ApJ...770...13W,2015ApJ...801..123W,2017arXiv171002212W,2015ApJ...805L...8B,2012ApJ...751L..25V,2016ApJ...816...37V,2017arXiv171201886V,2017Natur.545..457D,2018arXiv180102641D,2017arXiv171001251M}. 
The dashed horizontal line indicates the Milky Way value as a reference 
\citep[$\sim 3 \times 10^{-3}$,][]{2013ARA&A..51..105C}. 
Where necessary, TIR (8--1000 $\micron$) measurements were converted to FIR luminosity 
using $L_{\rm TIR} \simeq 1.3 L_{\rm FIR}$ \citep{2013ARA&A..51..105C}. 
The diagonal solid-blue line indicates our best-fit to the $z > 5.7$ quasars. 
Errors are only indicated for the HSC quasars to enhance the clarity of the figure. 
}
\label{fig6}
\end{figure}

The [C\,\emissiontype{II}]/FIR luminosity ratio can reflect physical conditions in the star-forming clouds. 
It has long been known that this ratio is more than an order of magnitude smaller in 
sources with high $L_{\rm FIR}$ than low 
\citep[e.g.,][]{1997ApJ...491L..27M,2003ApJ...594..758L,2008ApJS..178..280B,2010ApJ...724..957S,2011ApJ...728L...7G,2013ApJ...774...68D}. 
This trend has been extensively studied in the high-redshift universe as well \citep[e.g.,][]{2013ARA&A..51..105C}. 
Possible causes of the {\it [C\,\emissiontype{II}]-deficit in FIR-luminous objects} include 
AGN contamination to $L_{\rm FIR}$ \citep{2014ApJ...790...15S}, 
reduction of C$^+$ abundance due to AGN-heating \citep{2015A&A...580A...5L}, 
charging of dust grains \citep{1997ApJ...491L..27M}, 
saturation of the line flux due to high gas density \citep{1999ApJ...527..795K}, 
high dust opacity with respect to gas in dust-bounded region due to an increase of the average 
ionization parameter \citep{2009ApJ...701.1147A,2011ApJ...728L...7G}, 
and high gas surface density of individual cloud \citep[or higher molecular-to-atomic gas fraction,][]{2017MNRAS.467...50N}. 

Figure \ref{fig6} shows this [C\,\emissiontype{II}]-deficit in the FIR-luminous regime with a compilation of galaxies 
in both the nearby and high-redshift universes, including $z \gtrsim 6$ quasars. 
Those quasars also show a deficit with increasing $L_{\rm FIR}$ 
\citep[e.g.,][]{2016ApJ...816...37V,2013ApJ...773...44W}. 
However, the HSC quasars\footnote{We use $L_{\rm FIR}$($T_d$ = 35 K) in Table \ref{tbl2} as this case 
yields better agreement with [C\,\emissiontype{II}]-based SFR than that of $T_d$ = 47 K. 
The same $T_d$ is assigned for J2216-0016 as well to keep consistency, 
although its SFR with $T_d = 47$ K better agrees 
with that derived from the [C\,\emissiontype{II}] luminosity.}, 
as well as other less-luminous quasars 
\citep[i.e., CFHQS quasars,][]{2013ApJ...770...13W,2015ApJ...801..123W,2017arXiv171002212W}, 
exhibit $L_{\rm [C\,\emissiontype{II}]}$/$L_{\rm FIR}$ ratios comparable to 
or slightly higher ratios than the local LIRG-class objects at fixed $L_{\rm FIR}$ \citep{2013ApJ...774...68D}. 
In terms of EW$_{\rm [C\,\emissiontype{II}]}$, we also found that the HSC quasars have comparable 
values to local star-forming galaxies in Table \ref{tbl2}, 
while the luminous-end quasars tend to show lower 
EW$_{\rm [C\,\emissiontype{II}]}$ \citep{2016ApJ...816...37V}. 

The physical origin of the difference in $L_{\rm [C\,\emissiontype{II}]}$/$L_{\rm FIR}$ 
between those optically-luminous and -faint quasars is thus of interest. 
Both a higher AGN contribution to $L_{\rm FIR}$ and reduction of C$^+$ abundance 
due to too strong X-ray irradiation can reduce 
the ratio in luminous-end AGNs \citep{2014ApJ...790...15S,2015A&A...580A...5L}. 
As the spatial extent of [C\,\emissiontype{II}] or FIR continuum-emitting regions are 
comparable between the HSC quasars and optically luminous quasars (see Table \ref{tbl3}), 
a higher charge on dust grains due to enhanced star-forming activity \citep{1997ApJ...491L..27M} 
and higher gas surface density that shields ionizing radiation or cosmic rays 
could also cause the discrepancy as well \citep{2017MNRAS.467...50N}. 

High-redshift quasars follow a correlation in this plane 
that has comparable slope to the relation seen for lower redshift objects, 
but is offset to the higher $L_{\rm [C\,\emissiontype{II}]}$/$L_{\rm FIR}$ direction. 
We fit the relationship for the high-redshift quasar sample including the objects in this study 
following \citet{2017arXiv171002212W}. 
The orthogonal distance regression gives 
\begin{equation}
\log_{\rm 10} L_{\rm [C\,\emissiontype{II}]}/L_{\rm FIR} = (3.84 \pm 0.57) - (0.56 \pm 0.05) ~\log_{\rm 10} L_{\rm FIR}, 
\end{equation}
which is consistent with the relation derived in \citet{2017arXiv171002212W}. 
This relation is offset from the trend for the local galaxies and U/LIRGs by a factor of $\sim 2-3$ in the sense 
that $L_{\rm [C\,\emissiontype{II}]}$/$L_{\rm FIR}$ is higher for the quasars. 
It is also noteworthy that high-redshift FIR-luminous star-forming galaxies 
(i.e., non-AGN objects) also show an offset from the local trend \citep[e.g.,][]{2009A&A...500L...1M}. 
Thus, this trend is characteristic of high-redshift FIR-luminous objects 
regardless of the nature of their nuclear heating source(s). 
This could be due to the fact that those objects have more available gas \citep{2017MNRAS.467...50N}
: at a fixed $L_{\rm FIR}$ or SFR, 
higher gas mass can translate to lower gas mass {\it surface} density of individual cloud (linked to a star formation efficiency), 
which reduces the ability of shielding ionizing photons and cosmic rays. 
In that case, we would expect a higher [C\,\emissiontype{II}]/CO abundance ratio, 
as the column density for shielding ionizing photons and cosmic rays decreases there (low surface density). 
However, a more thorough study of the true shape 
of the FIR spectra of high-redshift objects 
is necessary to accurately constrain $L_{\rm [C\,\emissiontype{II}]}$/$L_{\rm FIR}$ ratios 
and to infer the physical properties of the gas content \citep[e.g.,][]{2010ApJ...724..957S} in the early universe.

\subsection{Cold dust content}
The rapid enrichment of the ISM at the $z \gtrsim 6$ universe, 
which is reflected in the large amounts of cold dust in quasar hosts 
\citep[e.g., several $\times$ 10$^8$ $M_\odot$ in SDSS quasars,][]{2008ApJ...687..848W,2011AJ....142..101W}, 
has been a challenge for many theoretical studies \citep[][for a recent review]{2017PASA...34...31V}. 
Similar high (or even higher) $M_{\rm dust}$ have also been observed in SMGs 
at $z > 6$ as well \citep{2013Natur.496..329R,2017ApJ...842L..15S}. 
Recent models particularly stress the importance of grain growth in cold, dense gas clouds 
rather than stellar yields as the dominant source of the early dust production 
\citep[e.g.,][]{2010A&A...522A..15M,2015MNRAS.451L..70M,2017MNRAS.471.3152P}. 
In these models, the observed $M_{\rm dust}$ of both optically/FIR-luminous quasars (e.g., SDSS) 
and faint HSC quasars (several $\times$ 10$^7$ $M_\odot$; Table \ref{tbl2}) can be reproduced 
by modifying the characteristic dust accretion time scale \citep[$\tau_{\rm acc}$, e.g.,][]{2014MNRAS.437.1636H}. 

According to the latest dust evolution model of \citet{2017MNRAS.471.3152P}, 
which incorporates several other dust formation and destruction processes as well, 
a $\tau_{\rm acc} \sim 15$ Myr model can yield $M_{\rm dust} \sim$ several $\times$ $10^8~M_\odot$ 
in a galaxy with a stellar mass ($M_*$) of several $\times$ 10$^{10}$ $M_\odot$ at $z \sim 6$. 
Meanwhile, decelerated dust growth with $\tau_{\rm acc} \sim 100$ Myr 
is sufficient to produce $M_{\rm dust}$ $\sim$ several $\times$ 10$^7$ $M_\odot$ in a similar $M_*$ galaxy. 
This $M_*$ would be valid for the $z \gtrsim 6$ quasar hosts observed so far at the rest-FIR, 
as long as their dynamical masses (subsection 4.3) mostly reflect their $M_*$. 
As this timescale depends on the molecular gas density of the system ($n_{\rm H2}$) as 
$\tau_{\rm acc} \propto n^{-1}_{\rm H2}$, 
the possible factor of $\sim 10$ difference in gas mass surface density 
between optically-luminous quasars 
and the HSC quasars at $z \gtrsim 6$ (subsection 3.3) can explain 
the difference in $\tau_{\rm acc}$ as well. 
Note that $\tau_{\rm acc}$ also depends on the inverse of the gas phase metallicity. 
Thus, if the difference in $M_{\rm dust}$ is eventually 
attributable to the difference in $n_{\rm H2}$, 
we expect that the gas-phase metallicity of optically-luminous 
and faint systems will not differ markedly. 
On the other hand, if $n_{\rm H2}$ is somehow comparable between them, 
the less luminous HSC quasars would have $\sim 1/10$ 
the gas metallicity as would the luminous-end quasars. 
These can be tested by a multi-species excitation analysis \citep[e.g.,][]{2016ApJ...830...53W} 
or by metallicity measurements that combine multiple fine-structure lines \citep[e.g.,][]{2012A&A...542L..34N}.

\subsection{Star-forming activity}
\begin{figure}[h] 
\begin{center}
\includegraphics[scale=0.2]{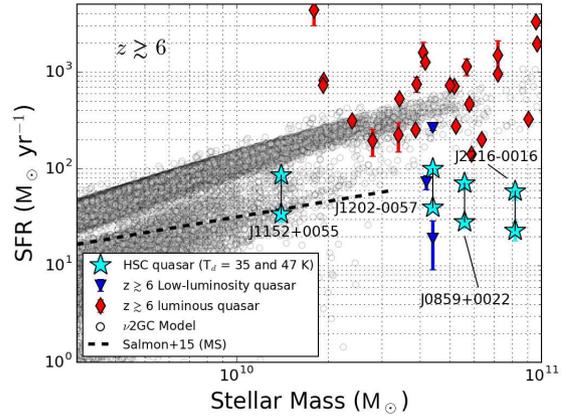}
\end{center}
\caption{
SFR$_{\rm TIR}$ as a function of $M_*$ for the HSC quasars (cyan stars), 
the $z \gtrsim 6$ luminous ($M_{\rm 1450} \lesssim -26$; red diamonds) 
and the $z \gtrsim 6$ low-luminosity ($M_{\rm 1450} > -25$; blue triangles) quasars 
having $10^{10} \leq M_{\rm dyn}/M_\odot \leq 10^{11}$. 
For the HSC quasars, we show both cases of $T_d$ = 35 K and 47 K, 
to exhibit the level of uncertainty due to unconstrained $T_d$. 
Those $M_{\rm dyn}$ are used as surrogates for $M_*$ in this plot. 
Background circles show simulated $z \sim 6$ galaxies 
hosting $M_{\rm BH} \geq 10^7~M_\odot$ SMBHs from the $\nu^2$GC model. 
Two sequences, namely the starburst-sequence and the star-forming main-sequence (MS), are apparent in the model. 
The latter is consistent with the recently suggested MS at $z \sim 6$ 
from rest-frame UV-to-NIR photometric observations \citep[][black-dashed line]{2015ApJ...799..183S}. 
}
\label{fig7}
\end{figure}

The majority of star-forming galaxies are found to populate the so-called {\it main sequence (MS)} of star formation 
on the $M_*$--SFR plane \citep[e.g.,][]{2007ApJ...670..156D,2007A&A...468...33E}, 
and the evolution of the MS over cosmic time has been studied extensively, 
even up to $z \sim 5-6$ \citep{2014ApJS..214...15S,2014ApJ...791L..25S,2015A&A...581A..54T,2015ApJ...799..183S}. 
The MS can be used to define starburst galaxies, normal star-forming galaxies, 
and quenched/quiescent galaxies at each redshift. 
While the MS is not well constrained at $z \gtrsim 5$, 
it is still informative to place $z \gtrsim 6$ quasars on the $M_*$ vs SFR plane 
and compare the levels of star formation among other galaxies. 

To this end, we first computed dynamical masses ($M_{\rm dyn}$) of the HSC quasars 
by following the standard procedure in previous $z \gtrsim 6$ quasar studies 
\citep[e.g.,][]{2010ApJ...714..699W,2013ApJ...773...44W,2015ApJ...801..123W,2016ApJ...816...37V}: 
here, we assumed that the observed [C\,\emissiontype{II}] emission came from a thin disk 
such that the velocity structure reflects rotational motion. 
The inclination angle of the disk ($i$) is determined by the ratio 
of the deconvolved (Table \ref{tbl3}) major ($a_{\rm maj}$) and minor ($a_{\rm min}$) 
axes of the [C\,\emissiontype{II}] emitting regions, 
$i$ = cos$^{-1}$($a_{\rm min}$/$a_{\rm maj}$). 
The circular velocity ($v_{\rm circ}$) is calculated as $v_{\rm circ}$ 
= 0.75 FWHM$_{\rm [C\,\emissiontype{II}]}$/$\sin i$ \citep{2010ApJ...714..699W}: 
the FWHM$_{\rm [C\,\emissiontype{II}]}$ of the HSC quasars are listed in Table \ref{tbl2}. 
The disk size (diameter, $D$) is approximated as $D$ = 1.5 $\times$ $a_{\rm maj}$ 
to account for the spatially extended component (i.e., full width at 20\% of the peak intensity for a Gaussian profile) 
to keep consistency with the previous works shown above. 

Then, the $M_{\rm dyn}$ enclosed within $D$ is given by, 
\begin{equation}
M_{\rm dyn}/M_\odot = 1.16 \times 10^5 \left(\frac{v_{\rm circ}}{\rm{km~s^{-1}}}\right)^2 \left(\frac{D}{\rm{kpc}}\right)
\end{equation}
The resultant values are listed in Table \ref{tbl4}. 
Note that the errors on $M_{\rm dyn}\sin^2i$ are estimated 
from the FWHM$_{\rm [C\,\emissiontype{II}]}$ and source size. 
On the other hand, formal errors on $M_{\rm dyn}$ themselves 
(i.e., after correcting for the inclination angles) 
are not given due to multiple uncertainties 
including those of the inclination angles and the true geometry of the line emitting regions. 

Keeping the existence of such large {\it unconstrained} systematic uncertainty in mind, 
hereafter we use the above $M_{\rm dyn}$ as a surrogate for $M_*$, 
which is a common procedure in high-redshift quasar studies 
\citep[e.g.,][]{2010ApJ...714..699W,2013ApJ...773...44W,2015ApJ...801..123W,2017arXiv171002212W,2016ApJ...816...37V}. 
Note that in three of our HSC quasars, gas masses obtained by applying a plausible gas-to-dust mass ratio 
\citep[e.g., 100,][]{2007ApJ...663..866D} to our derived $M_{\rm dust}$, are small relative to $M_{\rm dyn}$, 
indicating that stellar components dominate their $M_{\rm dyn}$. 
The one exception is J1152+0055, whose inferred gas mass is $\sim 50\%$ of $M_{\rm dyn}$, 
but this still has a limited impact on our conclusions. 
The resultant $M_*$ clearly constitute 
the massive end of $z \sim 6$ galaxies \citep[e.g.,][]{2015A&A...575A..96G}. 

\begin{table*}
  \tbl{Dynamical properties of the HSC quasars}
  {%
  \begin{tabular}{cccc} \hline \hline
   Name & $M_{\rm dyn} \sin^2i$ (10$^{10}$ $M_\odot$) & $M_{\rm dyn}$ (10$^{10}$ $M_\odot$) & $M_{\rm BH}$ (10$^8$ $M_\odot$) \\ \hline
   J0859+0022 & 3.3 $\pm$ 1.1 & 5.6 & 0.2$^{+0.2}_{-0.1}$ \\
   J1152+0055 & 1.1 $\pm$ 0.5 & 1.4 & 4.3$^{+7.8}_{-2.8}$ \\
   J2216-0016 & 6.4 $\pm$ 1.3 & 8.2 & 6.1$^{+6.1}_{-3.0}$ \\
   J1202-0057 & 2.9 $\pm$ 0.8 & 4.4 & 0.4$^{+0.7}_{-0.3}$ \\ \hline
  \end{tabular}}\label{tbl4}
  \begin{tabnote}
    Formal errors on $M_{\rm dyn}$ are not given due to multiple unconstrained uncertainties 
    including those of the inclination angles and the geometry of the line emitting regions. 
    See details about the $M_{\rm BH}$ measurement of J0859+0022 and J2216-0016 
    with the Mg\,\emissiontype{II}-based calibrations \citep{2009ApJ...699..800V} in M. Onoue et al. (in preparation). 
    For the remaining two quasars, Eddington-limited mass accretion is assumed, 
    which gives the lower limit on $M_{\rm BH}$. 
    A typical uncertainty is 0.3 dex for the Mg\,\emissiontype{II}-based $M_{\rm BH}$ \citep{2008ApJ...680..169S}, 
    and 0.45 dex for the Eddington ratio-based $M_{\rm BH}$ \citep{2015ApJ...801..123W}, respectively. 
  \end{tabnote}
  \end{table*}

The relationship between $M_*$ and SFR (calculated with $L_{\rm FIR}$) 
of the HSC quasars is plotted in Figure \ref{fig7}, 
along with $z \sim 6$ results from the $\nu^2$GC simulation: 
the simulated $M_*$ values (not $M_{\rm dyn}$) are used here. 
We selected $\simeq 41,000$ galaxies hosting $\geq10^7~M_\odot$ SMBHs 
from the simulated catalog (see also subsection 4.4), 
all of which are indeed high-$M_*$ galaxies. 
The simulated galaxies show two sequences, 
a starburst sequence (upper) and MS (lower): 
the gap between these two sequences is artificial 
due to the limited mass- and time-resolution 
of the model (see details in H. Shirakata et al. in preparation)
\footnote{We confirmed that this effect has little impact on time-integrated quantities 
such as $M_{\rm dyn}$, $M_*$, and $M_{\rm BH}$, which are used in subsection 4.4.}. 
Keeping this in mind, the simulated galaxies are used 
to infer the star-formation levels of the observed quasars. 
The model-MS is consistent with other semi-analytic models 
\citep[e.g.,][]{2008MNRAS.391..481S}, 
and roughly matches the recently {\it observed} MS at $z \sim 6$ 
\citep[investigated at $M_* \lesssim 10^{10.5}~M_\odot,$][]{2015ApJ...799..183S}. 
At lower-$z$, the model is also consistent with observations \citep[e.g.,][]{2007ApJ...670..156D,2007A&A...468...33E}. 
As seen in Figure \ref{fig7}, the HSC quasars studied here 
all reside {\it on} or even {\it below} the MS at $z \sim 6$, 
both compared to the $\nu^2$GC model and to the observed relationship. 
Therefore, we suggest that these HSC quasars are now 
ceasing their star-forming activities, and are transforming into the quiescent population. 

We also plot $z \gtrsim 6$ optically luminous 
($M_{\rm 1450} \lesssim -26$) and low-luminosity ($M_{\rm 1450} > -25$) 
quasars with [C\,\emissiontype{II}] observations 
\citep{2005A&A...440L..51M,2012ApJ...751L..25V,2016ApJ...816...37V,
2013ApJ...773...44W,2016ApJ...830...53W,
2015ApJ...805L...8B,2015ApJ...801..123W,2017arXiv171002212W,
2017Natur.545..457D,2018arXiv180102641D,2017arXiv171001251M} in Figure \ref{fig7}. 
Here, we limit the sample to have $10^{10} \leq M_{\rm dyn}/M_\odot \leq 10^{11}$ 
for fair comparison with the HSC quasars, 
after considering a recent argument that both the SFR 
and mass accretion rate may depend on the $M_*$ of the host galaxy \citep{2017ApJ...842...72Y}. 
Again their $M_{\rm dyn}$ are used as surrogates for $M_*$. 

We found that these optically-luminous quasars all reside {\it on} or {\it above} the MS, 
i.e., their hosts are indeed starburst galaxies. 
On the other hand, two of the three (optically) low-luminosity quasars 
\citep[CFHQS quasars,][]{2015ApJ...801..123W,2017arXiv171002212W} 
having comparable $M_{\rm 1450}$ to our HSC quasars, also exhibit similarly low SFR. 
Therefore, host galaxies of those optically luminous quasars and less luminous quasars 
including our HSC quasars, plotted in Figure \ref{fig7}, constitute different populations 
in terms of the evolutionary stages of star formation. 
The order of magnitude difference in gas mass surface density implied in this work 
could drive the above difference in both SFR 
and mass accretion rate \citep[e.g.,][]{2010MNRAS.407.1529H}. 
We also point out that a gap between the luminous quasars and the HSC quasars stands out. 
A transformation from the starburst phase to the quiescent phase would thus be quite rapid: 
merger-induced galaxy evolution models, for example, indeed predict such evolution 
\citep[e.g.,][]{2008ApJS..175..356H,2015MNRAS.449.1470V}. 

Note that, however, those FIR luminous quasars could only 
be a subset of all optically luminous quasars at $z \gtrsim 6$, 
according to the compilation in \citet{2014MNRAS.438.2765C}, 
as the majority of the quasars reported there only have upper limits on $L_{\rm FIR}$ 
(mostly measured with single-dish observations). 
The reported upper limits in \citet{2014MNRAS.438.2765C} 
are typically $\sim (3-6) \times~10^{12}~L_\odot$. 
Thus, it is plausible that a large number of optically luminous quasars 
could lie on the MS, which makes the actual fraction of $z \gtrsim 6$ luminous quasars 
that are hosted by starburst galaxies highly uncertain.   
Meanwhile, the recent ALMA survey toward $z \gtrsim 6$ optically luminous quasars, 
without prior information on their FIR fluxes, 
revealed that the bulk of them indeed have ULIRG-like $L_{\rm FIR}$ 
\citep[][]{2018arXiv180102641D}. 
It is therefore also possible that the $z \gtrsim 6$ quasars 
compiled in \citet{2014MNRAS.438.2765C} 
actually have ULIRG-like (or luminous LIRG-like) $L_{\rm FIR}$, 
which are yet below the detection limits of previous single dish observations. 
Deeper submm observations are clearly required to depict 
the true quasar distribution on the $M_*$--SFR plane. 

It is also noteworthy that the spatial extents (both [C\,\emissiontype{II}] and continuum) of the HSC quasars 
are consistent with the typical size of the stellar components of some 
compact quiescent galaxies \citep[cQGs, e.g.,][]{2008ApJ...677L...5V,2014ApJ...797...17K} 
exhibiting little ongoing star formation. 
Such cQGs have been found even at $z \sim 4$ \citep{2014ApJ...783L..14S,2015ApJ...808L..29S}. 
Multi-band photometric analysis \citep{2014ApJ...783L..14S} suggests 
that the stellar mass of $z \sim 4$ cQGs (several $\times$ 10$^{10}~M_\odot$) had been formed 
at $z \sim 6$ with intense starburst (characteristic time scale $\sim 100$ Myr). 
Within the context of this scenario, the host galaxies of HSC quasars at $z \sim 6$ 
may represent an earlier phase in the formation of massive compact galaxies, 
possible through mergers, when star formation is beginning to its cessation. 
Indeed, recent studies on the sizes and star formation histories 
of submillimeter galaxies (SMGs) and cQGs suggest 
that cQGs are descendants of SMGs at higher redshifts, 
having passes through the optical quasar phase, which will further evolve 
into local giant ellipticals through dry mergers \citep[e.g.,][]{2014ApJ...782...68T,2015ApJ...810..133I,2017ApJ...850...83F}.

\subsection{Early SMBH-host galaxy co-evolution in the less-luminous quasars}
Finally, we investigate the nature of the early co-evolution of SMBHs and their hosts. 
To this end, we compute $M_{\rm BH}$ of the HSC quasars (Table \ref{tbl4}). 
For J0859+0022 and J2216-0016, a virial calibration using broad Mg\,\emissiontype{II} emission line \citep{2009ApJ...699..800V} 
was applied to derive their $M_{\rm BH}$ (M. Onoue et al. in preparation). 
It is noteworthy that J0859+0022 
has a very low $M_{\rm BH}$ ($\sim 2 \times 10^7~M_\odot$) 
as compared to previously known $z \gtrsim 6$ quasars, 
clearly demonstrating the high sensitivity of our HSC survey to discover such lower mass objects. 
For the remaining two cases, the Eddington-limited mass accretion was assumed, 
following previous $z \gtrsim 6$ quasar studies \citep[e.g.,][]{2013ApJ...773...44W,2016ApJ...816...37V}, 
which gives the lower-limit on $M_{\rm BH}$. 
Their bolometric luminosities were calculated from the rest-frame UV luminosity 
at 1450 {\AA} with a correction factor of 4.4 \citep{2006ApJS..166..470R}. 
The $M_{\rm dyn}$ values derived in subsection 4.3 are again used here as surrogates for $M_*$. 

We have also compiled $M_{\rm BH}$ or $M_{\rm 1450}$ 
\citep{2003ApJ...587L..15W,2010AJ....140..546W,2014ApJ...790..145D,
2015ApJ...801L..11V,2015MNRAS.453.2259V,2015ApJ...798...28K,
2015Natur.518..512W,2016ApJS..227...11B,2017arXiv171201860B,
2016ApJ...833..222J,2017ApJ...845..138S,2017arXiv171001251M,2018arXiv180102641D} 
and [C\,\emissiontype{II}]-based $M_{\rm dyn}$ 
\citep{2005A&A...440L..51M,2012ApJ...751L..25V,2016ApJ...816...37V,2017arXiv171201886V,
2013ApJ...773...44W,2016ApJ...830...53W,2015ApJ...805L...8B,2015ApJ...801..123W,2017arXiv171002212W,
2017Natur.545..457D,2018arXiv180102641D,2017arXiv171001251M} 
measurements for $z \gtrsim 6$ quasars from the literature. 
The $M_{\rm BH}$ estimates are based on Mg\,\emissiontype{II} measurements 
with a typical uncertainty of 0.3 dex \citep{2008ApJ...680..169S}, 
which is added in quadrature to their measurement uncertainties. 
We assumed Eddington-limited accretion for those objects without Mg\,\emissiontype{II} data 
after deriving their bolometric luminosities from $M_{\rm 1450}$: 
0.45 dex uncertainty is assumed for those $M_{\rm BH}$ \citep{2013ApJ...773...44W,2015ApJ...801..123W}. 
While this provides a lower limit to their $M_{\rm BH}$, this assumption would be reasonable 
since optically luminous $z \gtrsim 6$ quasars are known to radiate closely at the Eddington limit \citep[e.g.,][]{2010AJ....140..546W}. 
Following the procedure of \citet{2017arXiv171002212W}, 
we assigned $D$ = 4.5 kpc and random orientation angle $i$ = 55$^\circ$ 
for some unresolved host galaxies to estimate their average sizes. 

\begin{figure*}[hptb] 
\begin{center}
\includegraphics[scale=0.35]{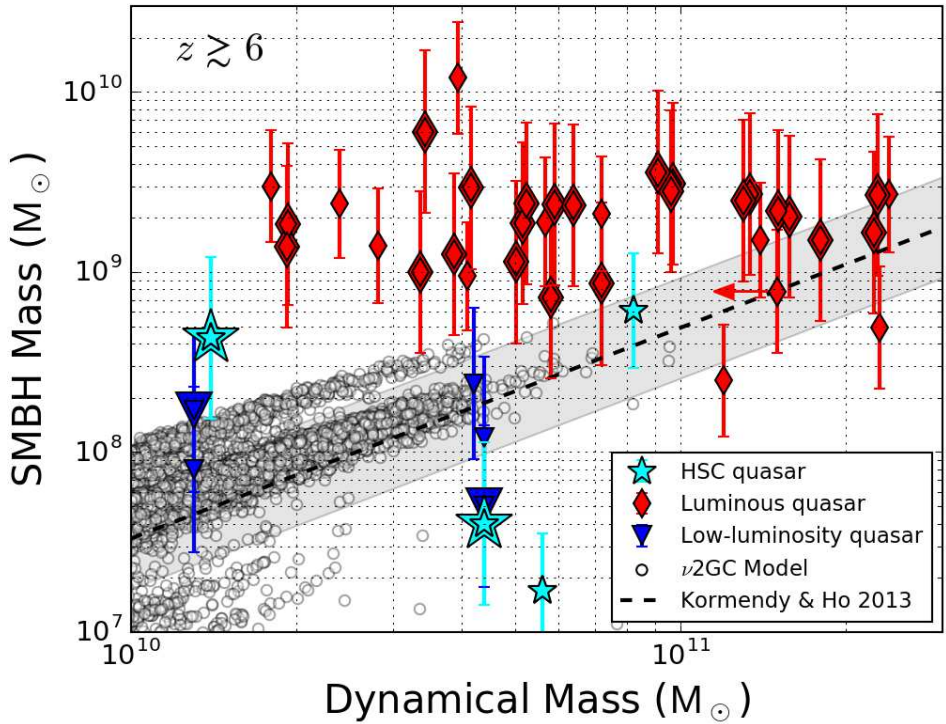}
\end{center}
\caption{
$M_{\rm BH}$ vs host galaxy $M_{\rm dyn}$ for $z \gtrsim 6$ quasars. 
The HSC quasars (cyan stars) are shown 
along with $\nu^2$GC model predictions at $z \sim 6$ (white circles). 
The diagonal dashed line indicates the local $M_{\rm BH}$--$M_{\rm bulge}$ 
relationship with its intrinsic scatter in the shaded region \citep{2013ARA&A..51..511K}. 
We equate $M_{\rm dyn}$ and $M_{\rm bulge}$ in this plot. 
Also shown are $z \gtrsim$ 6 optically luminous ($M_{\rm 1450} \lesssim -26$) quasars (red diamonds) 
and less-luminous ($M_{\rm 1450} > -25$; similar to the HSC quasars) ones (blue triangles). 
The less-luminous quasars, including the HSC ones, lie close to the local relation, 
whereas the luminous quasars show departures particularly at $M_{\rm dyn} < 10^{11}~M_\odot$. 
Among the four HSC quasars, Mg\,\emissiontype{II}-based $M_{\rm BH}$ 
is available for J0859+0022 and J2216-0016, 
whereas the Eddington-limited accretion is assumed for the rest (see Table \ref{tbl4}). 
The double symbols indicate that the Eddington limited accretion is assumed to derive their $M_{\rm BH}$. 
}
\label{fig8}
\end{figure*}

All of the above-mentioned quasars are plotted in Figure \ref{fig8}, 
overlaid with the local $M_{\rm BH}$--$M_{\rm bulge}$ relation \citep{2013ARA&A..51..511K}. 
The $M_{\rm dyn}$ of the central few kpc regions of high-redshift quasars 
can be equivalent to local $M_{\rm bulge}$ in 
the absence of further gas accretion and/or mergers. 
Thus, it is instructive to place the high-redshift quasars on this plane. 
As selection bias favoring the most luminous objects (or most massive objects) 
would have distorted the shape of the early co-evolutionary relations studied previously (section 1), 
we further divide the literature sample into 
(i) optically luminous objects ($M_{\rm 1450} < -25$; mostly $< -26$) 
and (ii) less luminous objects ($M_{\rm 1450} > -25$), 
based on the compiled quasar and galaxy luminosity functions at $z \sim 6$ in \citet{2016ApJ...828...26M}. 

Figure \ref{fig8} confirms the previous argument 
of \citet{2017arXiv171002212W} that there is 
no clear correlation between the two quantities 
when we focus on the all $z \gtrsim 6$ quasars observed so far, 
and the scatter is much larger than the local relation, 
particularly at $M_{\rm dyn} < 10^{11}~M_\odot$. 
At least in this galaxy mass range, the underlying $M_{\rm BH}$ distribution would have a wide scatter, 
and thus observations can be biased toward more luminous 
or more massive objects \citep[e.g.,][]{2007ApJ...670..249L,2014MNRAS.438.3422S}. 
This is demonstrated by the different distributions of 
the optically luminous (e.g., SDSS quasars) 
and less luminous (including HSC) quasar populations: 
the luminous quasars clearly have over-massive SMBHs as compared to the local relation in this mass range, 
whereas less luminous quasars are roughly consistent with the local relation within their uncertainties. 
We now see this difference more clearly than previous studies, 
since we almost double the number of the low luminosity quasars with $M_{\rm dyn}$ measurements in this study. 
Therefore, our study highlights the importance of probing low luminosity quasars 
to understand the unbiased early co-evolutionary relation 
reflecting the bulk of the AGN-host galaxies in this epoch, 
although the sample is still too small to statistically claim this argument. 

We also compare the observed distributions of $z \sim 6$ quasars on the plane 
with simulated galaxies from the $\nu^2$GC model, which are also plotted in Figure \ref{fig8}. 
Here, we selected all galaxies containing $M_{\rm BH} \geq 10^7~M_\odot$ 
SMBHs at $z \sim 6$ as we focused on massive quasars. 
The simulation traces the SMBH growth from the seed mass of $10^3~M_\odot$
\footnote{Changing the seed BH mass to $10^5~M_\odot$ does not 
affect the results at the high $M_{\rm BH}$ or high $M_{\rm bulge}$ regions, 
primarily because the $\nu^2$GC model allows super-Eddington accretion \citep{2016MNRAS.461.4389S}}. 
The galaxy bulge and the central SMBH gain masses, 
while maintaining the relation \citep{2016PASJ...68...25M}, 
\begin{equation}
\Delta M_{\rm BH} = f_{\rm BH} \Delta M_{*,\rm burst}, 
\end{equation}
where $\Delta M_{*,\rm burst}$ is the total mass of stars 
newly formed during a starburst episode in a {\it bulge} 
induced mainly by galaxy mergers, 
$\Delta M_{\rm BH}$ is the total SMBH mass growth, 
and $f_{\rm BH}$ is a constant (= 0.01) selected 
to match the local $M_{\rm BH}$--$M_{\rm bulge}$ relation. 
It is thus apparent that the simulated galaxies tend to follow the local relation. 
Note that we use $M_{\rm dyn}$ returned by the model, 
which indicates either (i) total mass within a half-mass radius (bulge-dominated galaxy) 
or (ii) total mass within a disk effective radius (disk-dominated galaxy). 
The mass of the dark matter would not be important at these spatial scales \citep{2017Natur.543..397G}. 
Thus, our comparison with observed data is fair. 

It is intriguing in Figure \ref{fig8} that $M_{\rm BH}$ and $M_{\rm dyn}$ 
of the low luminosity quasars (HSC + CFHQS) are close to the simulated values. 
This implies, based on the $\nu^2$GC model, 
that these quasars could have been formed through 
the standard, (quasi-) synchronized galaxy--SMBH formation scenario 
\citep[e.g.,][]{2005Natur.433..604D,2006ApJS..163....1H}, 
although we cannot exclude other evolutionary scenarios. 
The significantly lower $M_{\rm BH}$ of J0859+0022 ($\sim 2 \times 10^7~M_\odot$) 
than the local relation seems to support that standard galaxy evolutionary scheme, 
where a starburst phase (or growth of stellar content) occurs earlier 
and an SMBH growth later \citep[see also a recent ALMA work by][]{2017arXiv171203350U}. 
Note that expected halo masses are $\sim$ several $\times~10^{12}~M_\odot$ 
for those lower mass quasars based on our model, 
which will be observationally tested in future. 

On the other hand, it is still challenging to form 
massive-end {\it galaxies} ($M_{\rm dyn} \gtrsim 10^{11}~M_\odot$) 
that contain $M_{\rm BH} \gtrsim 5 \times 10^8~M_\odot$ 
SMBHs with this $\nu^2$GC simulation. 
We would also point out that the scatter 
around the local relation is smaller at $M_{\rm dyn} \gtrsim 10^{11}~M_\odot$, 
likely indicating (several episodes of) AGN feedbacks to regulate galaxy growth 
to finally converge to the local relation. 
As such massive-end objects should be quite rare, 
this remains a room that we will be able to generate those objects 
once we simulate much larger volumes. 
It is, however, virtually impossible to form quasars having 
$M_{\rm dyn} \lesssim 10^{11}~M_\odot$ 
and $M_{\rm BH} \gtrsim 5 \times 10^8~M_\odot$ with our model, 
as long as we use the fixed $f_{\rm BH}$. 
One other possibility is that these over-massive objects were formed through 
different path(s) from those incorporated in the $\nu^2$GC simulation. 
For example, additional supply of cold gas directly 
from the intergalactic medium would boost the mass, 
especially in the nuclear region of galaxies \citep[e.g.,][]{2009Natur.457..451D}: 
the potential importance of fueling mechanisms 
other than the standard merger picture has been 
investigated recently \citep[e.g.,][]{2017ApJ...836....8T}. 
It is even plausible that such quasars with over-massive SMBHs 
will accordingly evolve into {\it galaxies with over-massive SMBHs} at $z \sim 0$, 
which are start to be found recently \citep{2012Natur.491..729V}, 
although dry mergers will move them toward the local relation as time goes by from $z \sim 6$ to $\sim 0$. 
Further investigations of galaxy properties as well as 
environments around the quasars are essential to 
better understand the underlying processes of co-evolution.

\section{Summary and future prospects}\label{sec5}
We have presented ALMA observations of four optically low-luminosity 
($M_{\rm 1450} > -25$) quasars at $z \gtrsim 6$ recently discovered 
by our wide and deep optical survey with the Hyper Suprime-Cam (HSC) 
on the Subaru telescope \citep{2016ApJ...828...26M}. 
This study significantly increased the known sample in the low-luminosity regime at $z \gtrsim 6$ 
reported in works of \citet{2013ApJ...770...13W,2015ApJ...801..123W,2017arXiv171002212W}, 
giving us a less-biased view of high-redshift galaxy-SMBH evolution.  
All four quasars have been detected in the [C\,\emissiontype{II}] emission line 
and the underlying rest-FIR continuum emission. 
The main findings of this paper can be summarized as follows: 

\begin{itemize}
\item[1.] The [C\,\emissiontype{II}] line fluxes are as low as 0.4--1.1 Jy km s$^{-1}$, 
which corresponds to line luminosities of $L_{\rm [C\,\emissiontype{II}]} \simeq (4-10) \times 10^8~L_\odot$. 
These are more than one order of magnitude fainter 
than in $z \gtrsim 6$ optically-luminous 
quasars \citep[e.g.,][]{2013ApJ...773...44W,2016ApJ...816...37V}, 
and are comparable to local LIRGs \citep{2013ApJ...774...68D}. 
The line FWHM ranges from 192 to 356 km s$^{-1}$, similar 
to other $z \gtrsim 6$ quasars \citep[e.g.,][]{2013ApJ...773...44W,2015ApJ...801..123W,2016ApJ...816...37V}. 
The inferred star formation rates are similarly low (SFR $\sim 25-67~M_\odot$ yr$^{-1}$) 
as compared to the luminous quasar hosts ($>100~M_\odot$ yr$^{-1}$). 
\item[2.] The underlying rest-FIR continuum emission was 
detected in all objects (136--246 $\mu$Jy beam$^{-1}$). 
A modified black body fit with a dust temperature $T_d$ = 47 K 
to the measurements yielded FIR luminosities 
$L_{\rm FIR} \simeq (3-5) \times 10^{11}~L_\odot$, i.e., LIRG-class $L_{\rm FIR}$. 
Meanwhile, we found that $T_d$ = 35 K ($L_{\rm FIR} \simeq (1-2) \times 10^{11}~L_\odot$) 
yields IR-based SFR ($23-40~M_\odot$ yr$^{-1}$) 
that better match the [C\,\emissiontype{II}]-based SFR. 
In either case, the inferred dust masses are several $\times$ 10$^7$ $M_\odot$, 
which are an order of magnitude smaller than 
the optically-luminous quasar host galaxies at $z \gtrsim$ 6 
\citep[e.g.,][]{2007AJ....134..617W,2016ApJ...816...37V}. 
\item[3.] The spatial extent of the barely resolved [C\,\emissiontype{II}] emitting regions 
are mostly $\sim 3$ kpc for the HSC quasars (J2216-0016 shows $\sim 5$ kpc), 
which agree with the continuum-derived sizes despite the large errors. 
These numbers are comparable to those of the optically-luminous quasars 
having at least an order of magnitude higher $L_{\rm FIR}$ (or dust mass). 
It thus implies that the correspondingly different ISM mass surface density \citep{2010MNRAS.407.1529H} 
drives the difference in AGN activity between the HSC quasars and the more luminous quasars. 
\item[4.] We did not find any continuum or line emitter 
physically close to the HSC quasars within the nominal FoVs, 
except for one likely lower-redshift weak line emitter. 
Recent number counts suggest that we could have seen some objects 
in our deep observations (5$\sigma$ = 44--104 $\mu$Jy beam$^{-1}$), 
but several factors could reconcile this discrepancy. 
\item[5.] The BAL quasar J2216-0016 seems to show two components 
in the [C\,\emissiontype{II}] emission line spectrum and its velocity-integrated spatial distribution. 
This may reflect either [C\,\emissiontype{II}]-outflows or a galaxy merger. 
Higher resolution observations are required to further elucidate the nature of this high-redshift BAL quasar host. 
\item[6.] The $L_{\rm [C\,\emissiontype{II}]}/L_{\rm FIR}$ ratios of the HSC quasars 
are fully consistent with the local LIRG-class objects, whereas optically-luminous quasars 
tend to show a [C\,\emissiontype{II}]-deficit trend at increasing $L_{\rm FIR}$. 
This suggests that a star formation mode similar to local LIRGs 
(not ULIRG-like bursts) prevails in the HSC quasars. 
The order of magnitude of difference in the SFR (and likely ISM mass) surface densities 
between the HSC quasars and optically-luminous quasars may be the one of the physical origin(s) of the deficit. 
\item[7.] Our attempt to place the HSC quasars on the stellar mass 
($M_*$; we used dynamical masses $M_{\rm dyn}$ as surrogates for them) vs. SFR plane 
suggests that the HSC quasars and other similarly less-luminous quasars 
are on or even below the $z \sim 6$ star formation main sequence (MS), 
i.e., they are now ceasing their star formation. 
This is supported by both recent observations 
and a semi-analytical galaxy evolution model 
\citep[$\nu^2$GC,][]{2016PASJ...68...25M}. 
As optically-luminous quasars reside on or even above the MS (i.e., starburst galaxies), 
there could be an evolutionary difference between these luminous and less luminous quasar hosts. 
\item[8.] Our dynamical measurements suggest that the HSC quasars 
along with similarly less luminous quasars at $z \gtrsim 6$ \citep[e.g.,][]{2015ApJ...801..123W,2017arXiv171002212W}  
tend to follow the local co-evolutionary relation, whereas luminous objects 
show clear departures from it (over-massive SMBHs) particularly at $M_{\rm dyn} < 10^{11}~M_\odot$. 
This highlights the importance of probing less luminous quasars 
to depict the unbiased shape of the early co-evolution. 
The mass properties of those less-luminous quasars 
can be reproduced by the $\nu^2$GC model, 
implying they could be formed with quasi-synchronized galaxy (bulge)--SMBH evolution scheme, 
although we do not argue that this is the only scenario to explain the results. 
On the other hand, we may need to consider some other evolution paths 
to generate galaxies hosting over-massive SMBHs. 
\end{itemize}

A higher spatial resolution that that provided in this study ($\sim 0\arcsec.5$), 
which is achievable with ALMA, is necessary to elucidate 
the physical origin of the spectral peculiarity of J2216-0016. 
Furthermore, the trends of low-luminosity quasars shown above, 
which are clearly different from those of optically luminous quasars, 
are based on the small sample. 
This will be statistically confirmed with our growing SHELLQs sample.

\bigskip
\begin{ack}
We thank the anonymous referee for his/her thorough reading and useful comments which greatly improved this paper. 
T.I. particularly thank Dr. T. Suzuki at NAOJ for her fruitful comments. 
This paper makes use of the following ALMA data: ADS/JAO.ALMA\#2016.1.01423.S. 
ALMA is a partnership of ESO (representing its member states), NSF (USA) and NINS (Japan), 
together with NRC (Canada), NSC and ASIAA (Taiwan), and KASI (Republic of Korea), in cooperation with the Republic of Chile. 
The Joint ALMA Observatory is operated by ESO, AUI/ NRAO and NAOJ. 

The Hyper Suprime-Cam (HSC) collaboration includes the astronomical 
communities of Japan and Taiwan, and Princeton University. 
The HSC instrumentation and software were developed by the National Astronomical Observatory of Japan (NAOJ), 
the Kavli Institute for the Physics and Mathematics of the Universe (Kavli IPMU), 
the University of Tokyo, the High Energy Accelerator Research Organization (KEK), 
the Academia Sinica Institute for Astronomy and Astrophysics in Taiwan (ASIAA), and Princeton University. 
Funding was contributed by the FIRST program from Japanese Cabinet Office, 
the Ministry of Education, Culture, Sports, Science and Technology (MEXT), 
the Japan Society for the Promotion of Science (JSPS), Japan Science and Technology Agency (JST), 
the Toray Science Foundation, NAOJ, Kavli IPMU, KEK, ASIAA, and Princeton University. 

This paper makes use of software developed for the Large Synoptic Survey Telescope. 
We thank the LSST Project for making their code available as free software at http://dm.lsstcorp.org. 

The Pan-STARRS1 Surveys (PS1) have been made possible 
through contributions of the Institute for Astronomy, 
the University of Hawaii, the Pan-STARRS Project Office, 
the Max-Planck Society and its participating institutes, 
the Max Planck Institute for Astronomy, Heidelberg and the Max Planck Institute for Extraterrestrial Physics, Garching, 
The Johns Hopkins University, Durham University, the University of Edinburgh, Queen’s University Belfast, 
the Harvard-Smithsonian Center for Astrophysics, the Las Cumbres Observatory Global Telescope Network Incorporated, 
the National Central University of Taiwan, the Space Telescope Science Institute, 
the National Aeronautics and Space Administration under Grant No. NNX08AR22G 
issued through the Planetary Science Division of the NASA Science Mission Directorate, 
the National Science Foundation under Grant No. AST-1238877, 
the University of Maryland, and Eotvos Lorand University (ELTE). 

T.I., M.O., K.K, N.K., H.U., R.M., and T.N. are supported by JSPS KAKENHI grant numbers 
17K14247 (T.I.), 
15J02115 (M.O.), 
17H06130 (K.K.), 
15H03645 (N.K.), 
17K14252 (H.U.), 
15H05896 (R.M.), 
16H01101, 16H03958, and 17H01114 (T.N.), respectively. 
H.S. has been supported by the Sasakawa Scientific Research Grant from the Japan Science Society (29-214). 
T.I. is supported by the ALMA Japan Research Grant of NAOJ Chile Observatory, NAOJ-ALMA-175.
\end{ack}

\end{document}